\documentclass[]{pasj02}

\jyear{2025}
\Received{$\langle$reception date$\rangle$}
\Accepted{$\langle$acception date$\rangle$}
\Published{$\langle$publication date$\rangle$}

\usepackage{comment}
\usepackage{natbib,bm}

\usepackage[switch,mathlines]{lineno} 

\newcommand{\beq}{\begin{equation}}
\newcommand{\eeq}{\end{equation}}
\newcommand{\beqn}{\begin{eqnarray}}
\newcommand{\eeqn}{\end{eqnarray}}

\newcommand{\eqref}[1]{(\ref{#1})}

\graphicspath{{./}{figure/}}

\title{Nitrogen transport in protoplanetary disks by ammonium salts: a possible origin of Jupiter's nitrogen enrichment}
\author{Kanon \textsc{Nakazawa}\altaffilmark{1}}
\altaffiltext{1}{Department of Earth and Planetary Sciences, Institute of Science Tokyo, Meguro, Tokyo 152-8551, Japan}

\author{Satoshi \textsc{Okuzumi}\altaffilmark{1}}

\email{nakazawa.k.ai@m.titech.ac.jp}

\KeyWords{protoplanetary disks --- planets and satellites: composition} 

\maketitle

\begin{document}

\begin{abstract}
Atmospheric compositions preserve the history of planet formation processes. 
Jupiter has the remarkable feature of being uniformly enriched in various elements compared to the Sun, including highly volatile elements such as nitrogen and noble gases.
Radial transport of volatile species by amorphous ice in the solar nebula is one mechanism that explains Jupiter's volatile enrichment, but the low entrapment efficiency of nitrogen into amorphous ice is an issue.
We propose an alternative mechanism of delivering nitrogen to Jupiter: radial transport of semi-volatile ammonium salts in the solar nebula.
Ammonium salts have been identified in 67P/Churyumov-Gerasimenko and can potentially compensate for the comet's nitrogen depletion compared to the Sun.
We simulate the radial transport and dissociation of ammonium salts carried by dust in a protoplanetary disk, followed by the accretion of the gas and NH$_3$ vapor by a protoplanet, as well as the delivery of nitrogen to the planetary atmosphere from the salt-containing planetary core that undergoes dilution.
We find that when the dust contains 10--30 wt\% ammonium salts, the production of NH$_3$ vapor in the inner disk ($\sim 3$ au) by dissociated salts and the incorporation of the salt-derived NH$_3$ through core formation and subsequent gas accretion by the protoplanet result in a planetary nitrogen enrichment consistent with the observations of Jupiter.
Ammonium salts may thus play a vital role in developing the atmospheric composition of planets forming in the inner disk. 
Combining our model with future observations of the bulk compositions and isotopes of comets and other primordial bodies will help to further elucidate the elemental transport to the gas giants and ice giants in the solar system.    
\end{abstract}

\section{Introduction}
\label{sec:intro}
Planetary atmospheric compositions preserve the history of planet formation processes and elemental transport in protoplanetary disks.
Jupiter is the most well-studied gas planet for atmospheric composition by in situ observations with the Galileo probe, Cassini, and Juno spacecraft.
A remarkable feature of Jupiter's composition is that the ratios of O, C, N, S, P, Ar, Kr, and Xe to H are uniformly enriched by a factor of 2--4 compared to the protosolar values \citep{2004Icar..171..153W, 2009Icar..202..543F, 2017Sci...356..821B, 2020NatAs...4..609L}.
It is striking that even N and Ar, which are generally thought to be volatile, are equally enriched.

There are broadly two scenarios that can potentially explain Jupiter's elemental enrichment.
The first scenario invokes the accumulation of solids (pebbles or planetesimals) that were uniformly enriched in different elements into Jupiter's core, followed by 
the dissolution of the core into the atmosphere during Jupiter's internal evolution. This scenario is in line with the observations from Juno that indicate that Jupiter has a diluted core, implying that heavy elements in the core were transported to the atmosphere (e.g., \citealt{2018A&A...610L..14V, 2019ApJ...872..100D}; for a review, \citealt{ 2022Icar..37814937H}).
Furthermore, simulations of Jupiter formation show that heavy elements tend to dissolve into the envelope formed during core accretion  \citep{2007Icar..187..600I, 2011MNRAS.416.1419H, 2018A&A...611A..65B, 2020ApJ...900..133V}. 
Additionally, convective mixing acts to erase compositional gradients present in Jupiter's upper atmosphere, thereby leading to a more homogeneous atmospheric composition \citep{2018A&A...610L..14V, 2020A&A...638A.121M}.
\citet{2019AJ....158..194O} suggested that uniform enrichment can be explained if the Jovian core formed far beyond the Ar snow line ($> 30$ au), where Ar freezes on the dust. Similarly, \citet{2019A&A...632L..11B} proposed Jupiter formation near the N$_2$ snow line.
However, theoretical studies may raise issues in terms of planetary growth.
The mass of a gas giant core that begins migrating from $r > 10$ au would exceed Jupiter's mass before it reaches current Jupiter's orbit \citep{2016ApJ...823...48T, 2020ApJ...891..143T}.
As an alternative idea, \citet{2021A&A...651L...2O} proposed core formation in the cryogenic shadowed region produced by dust piled up near the H$_2$O snow line.
This approach does not require large-scale orbital migration, but it is uncertain whether dust can accumulate in the shadowed region and form Jupiter's core.

The second scenario considers radial transport of volatile elements by clathrate or amorphous ice that entraps volatile species \citep[e.g.,][]{2001ApJ...550L.227G, 2015ApJ...798....9M}.
\citet{2015ApJ...798....9M} proposed a model in which water molecules photo-desorbed from icy grains entrap volatile species when they recondense as amorphous ice in the outer disk regions.
This icy dust then transports volatile elements to the inner region of the disk through radial drift.
\citet{2019ApJ...875....9M} numerically simulated the model of \citet{2015ApJ...798....9M} and showed that uniform elemental enrichment of Jupiter is achievable if Jupiter forms near the amorphous-crystalline phase transition line (143K).
The advantage of this scenario is that it does not presume a cryogenic environment or large-scale planetary migration, but it has a drawback in nitrogen transport.
\citet{2007Icar..190..655B} show that N$_2$ has one orders of magnitude lower entrapment efficiency into amorphous ice than CO$_2$.
We review previous studies on volatile entrapment by amorphous ice in Appendix \ref{append:entrapment}.

The nitrogen composition of minor bodies in the solar system, which serves as an indicator of the nitrogen content in primordial disk dust, is significantly depleted compared to the solar composition.
Asteroids in the main belt are highly depleted in nitrogen, with N/Si on average three orders of magnitude lower than the Sun.
Even primitive meteorites such as CI chondrites show two orders of magnitude lower N/Si than the Sun, including Ryugu samples that  experienced little alteration on Earth \citep{2014prpl.conf..363P, 2021PhR...893....1O, 2023Sci...379.0431O}.
Comets, which exhibit the most primordial solid composition in the solar system, retain volatile nitrogen species that rarely remain in these meteorites, with NH$_3$ at 1.0 wt\% to water, HCN at 0.5 wt\%, and other volatile nitrogen species of the order of 0.01 wt\% \citep{2011ARA&A..49..471M}.
Despite the high primordiality of comets, N/Si is one order of magnitude lower than that of the Sun \citep{2021PhR...893....1O}.
This means that even if dust consisting of cometary composition with volatile nitrogen species were transported to the inner disk, it would be difficult to elevate the nitrogen abundance in the disk to a super-solar level.
The carriers of the missing nitrogen are still debated.

Potential carriers of the missing nitrogen are semi-volatile ammonium salts.
Recently, ammonium salts have been identified in 67P/Churyumov-Gerasimenko as evidence of buried nitrogen in comets \citep[detected NH$_4$HCO$_2$,][]{2020NatAs...4..533A, 2020Sci...367.7462P}.
\citet{2022MNRAS.516.3900A} also reported the detection of NH$_4$SH.
Ammonium salts are the salts of NH$_4^+$ formed by the protonation of NH$_3$ and are less volatile than NH$_3$.
Ammonium formate NH$_4$HCO$_2$ identified in cometary nuclei has a dissociation temperature of around 200 K and can transport nitrogen in the solid state to the vicinity of the H$_2$O snow line \citep{2016ApJ...829...85B, 2020Sci...367.7462P}.
Ammonium salts have various compositions corresponding to acids in interstellar space, and NH$_4$CN, NH$_4$[H$_2$NCO$_2$] would exist in comets.
Laboratory experiments simulating the space environment have shown that the dissociation temperature of these salts ranges from 150 to 250 K \citep{2011A&A...535A..47D, 2019ApJ...878L..20P}.
Continued detection of ammonium salts suggests that the comet may have abundant nitrogen reserves in the form of semi-volatiles such as salts.
\citet{2020Sci...367.7462P} pointed out that ammonium salts may be present in comets at 10--30 wt\% in dust as a nitrogen carrier to compensate for the comet's nitrogen deficiency relative to the sun.
In addition to comet exploration, observations of molecular clouds by James Webb Space Telescope have shown that NH$_3$ and NH$_4^+$ have been present since the beginning of star formation, with each having a mass abundance of 5-10 wt\% relative to water \citep{2023NatAs...7..431M}.

In this paper, we propose a new mechanism of enhancing nitrogen abundance at the inner region of the disk and Jupiter: transport of nitrogen with ammonium salts by the inward drifting dust.
Conventional disk models do not consider ammonium salts, and discussions of Jupiter's nitrogen abundance have focused only on chemical species which sublimate farther out than Jupiter's current orbit such as N$_2$ and NH$_3$ \citep[e.g.,][]{2019MNRAS.487.3998B, 2019A&A...632L..11B}.
Here, we calculate the radial transport of the gas and dust including ammonium salts and test the possibility of Jupiter's nitrogen enrichment by the salts.

This paper is organized as follows. We describe the models of the protoplanetary disk, dust and gas evolution, and evolution of planetary nitrogen abundance in Section \ref{sec:model}. We present the results of our numeric calculation in Section \ref{sec:results}. We give discussions in Section \ref{sec:discussion} and summarize in Section \ref{sec:conclusion}.

\section{Model}
\label{sec:model}

\subsection{Overview}
\label{subsec:model_overview}

\begin{figure}[t]
\includegraphics[width=\hsize, bb=0 0 361 376]{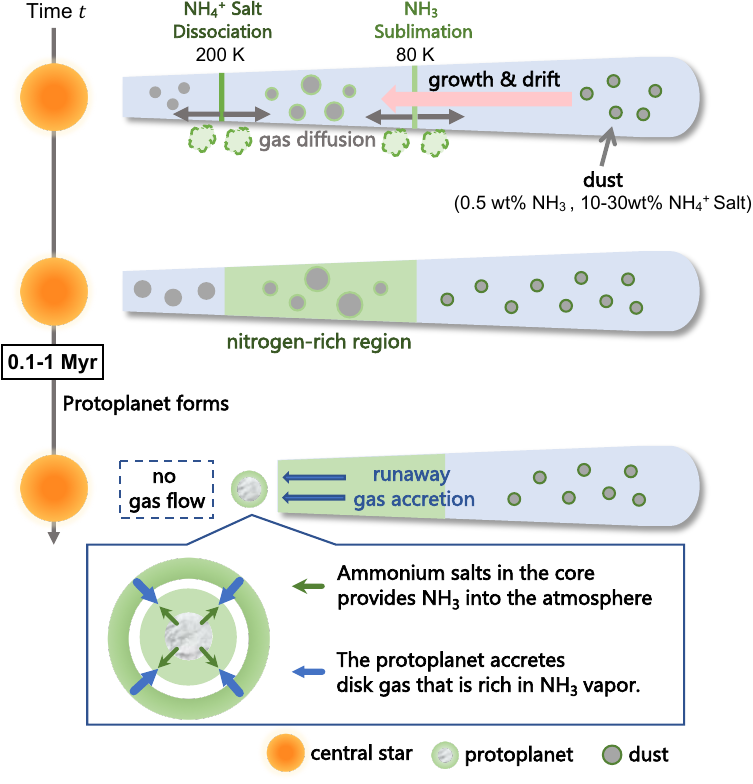}
\caption{Schematic illustration of our model. We consider a protoplanetary disk consisting of gas and dust. The dust grains contain ammonium salts and NH$_3$ ice as nitrogen carriers. The grains grow through mutual collisions and drift inward. When the grains crosses the sublimation/dissociation line, they release NH$_3$ vapor. A protoplanet forms at given time and orbit. This protoplanet contains ammonium salts in its core, and core erosion releases NH$_3$ into its atmosphere. The planet also accretes NH$_3$ vapor-rich disk gas during the gas accretion phase. {Alt text: Color illustration of our model.}}
\label{fig:conceptual_diagram}
\end{figure}

In this section, we describe our model for quantifying the contribution of ammonium salts to the nitrogen enrichment of a gas giant.
Figure \ref{fig:conceptual_diagram} presents a schematic of the model. We consider a protoplanetary disk consisting of gas and dust and assume that the dust contains semi-volatile ammonium salts in addition to volatile NH$_3$ ice. Accounting for ammonia sequestered in salts brings the nitrogen abundance of dust in the outer disk closer to the solar value. The model includes the evolution of the gas surface density, dust surface density, dust grain size, and NH$_3$ vapor surface density released from the dust.
The gas disk is evolved using the classical viscous disk model \citep{1974MNRAS.168..603L} (Section \ref{subsec:model_disk}).
The grains are initially 0.1 $\mu m$ in radius and are allowed to grow or fragment through mutual collisions (Section \ref{subsec:model_dust}).
The grains drift toward the central star due to radial drift caused by the friction arising from the sub-Keplerian rotation of the gas \citep{1972fpp..conf..211W, 1976PThPh..56.1756A, 1977MNRAS.180...57W}.
The inward-drifting grains release NH$_3$ vapor when they cross the orbits at which the temperature reaches the NH$_3$ ice sublimation and salt dissociation temperatures (assumed to be 200 K and 80 K, respectively). Assuming that these vapors are converted to highly volatile chemical species, we neglect recondensation of NH$_3$ (Section \ref{subsec:model_N_species}).
The released vapor enriches the inner disk region ($r \lesssim 10$ au) in nitrogen.
A planetary core forms from dust  containing ammonium salts and starts accreting the disk gas at given time and at fixed orbital radius.
We assume that the nitrogen incorporated into the core in the form of ammonium salts is delivered to the atmosphere by core dilution and subsequent convection in the atmosphere.
We calculate the nitrogen abundance of the growing planetary atmosphere until the planetary mass reaches current Jupiter's mass (Section \ref{subsec:model_planet}).

\subsection{Disk model}
\label{subsec:model_disk}
We adopt a viscous disk model around a solar-mass star $M_{\odot}$ \citep{1974MNRAS.168..603L}. 
The gas surface density $\Sigma_{\rm g}$ is evolved by the following equation,
\begin{equation}
\label{Sigma_g_evolution}
   \frac{\partial \Sigma_{\rm g}}{\partial t} = -\frac{1}{r} \frac{\partial}{\partial r}(rv_{\rm g}\Sigma_{\rm g}),   
\end{equation}
with
\begin{equation}
\label{v_g}
    v_{\rm g} = -\frac{3\nu_{\rm visc}}{r} \frac{\partial \ln{(r^{1/2} \nu_{\rm visc} \Sigma_{\rm g})}}{\partial \ln{r}},
\end{equation}
where $t$ is time, $r$ is the distance from the central star, $\nu_{\rm visc}$ is the gas viscosity. The gas viscosity is written as $\nu_{\rm visc} = \alpha c_{\rm s}^2/\Omega_{\rm K}$ \citep{1973A&A....24..337S}. Here, $\alpha$ is the dimensionless parameter controlling turbulence strength, $\Omega_{\rm K} = \sqrt{GM_{\odot}/r^3}$ is the Keplerian frequency with gravitational constant $G$, and $c_{\rm s} = \sqrt{k_{\rm B}T/\mu m_{\rm H}}$ is the isothermal sound speed with Boltzmann constant $k_{\rm B}$, mean molecular weight $\mu$ (taken to be 2.34), and proton mass $m_{\rm H}$. We assume a moderate level of turbulence and taken $\alpha$ to be either $10^{-3}$ or $10^{-4}$.
For simplicity, we assume that $\alpha$ is constant in time and space.
The initial gas surface density is given by \citep{1974MNRAS.168..603L, 1998ApJ...495..385H}
\begin{equation}
    \Sigma_{\rm g, 0} = \frac{M_{\rm disk}}{2\pi r_{\rm c}^2} \left(\frac{r}{r_{\rm c}}\right) \exp{\left(-\frac{r}{r_{\rm c}}\right)}, 
\end{equation}
where $M_{\rm disk}$ is the total mass of the disk and $r_{\rm c}$ is the characteristic radius of the disk. We assume a compact disk with $M_{\rm disk} = 0.05 M_{\odot}$ and $r_{\rm c} = 50~\rm{au}$. 
This value represents a typical disk mass and is also consistent with the gas mass estimated by the minimum-mass solar nebula model within a factor of three \citep{1981PThPS..70...35H}.

We consider stellar irradiation for optically thick disks and viscous accretion heating and give the disk temperature $T$ as \citep{1994ApJ...421..640N}
\begin{equation}
\label{T}
    T = (T^4_{\rm irr} + T^4_{\rm acc})^{1/4},
\end{equation}
where $T_{\rm irr}$ and $T_{\rm acc}$ are the contributions from irradiation and accretion heating, respectively.
The irradiation part is approximately given by \citet{1970PThPh..44.1580K} and \citet{1997ApJ...490..368C} as
\begin{equation}
\label{Tirr}
    T_{\rm irr} = 110 \left( \frac{r}{1 \rm au }\right)^{-3/7} \left( \frac{L}{L_{\odot} }\right)^{2/7} \left( \frac{M}{M_{\odot}}\right)^{-1/7} {\rm K},
\end{equation}
where $L$ and $M$ are the stellar luminosity and stellar mass, respectively.
We allow the stellar luminosity to evolve following equation (8) of \citet{2023ApJ...949..119K}, which is an empirical fit to the stellar evolution
model of \citet[see also \cite{2021ApJ...916...72M}]{2016A&A...593A..99F}.

The accretion part is written as \citep{1990ApJ...351..632H, 1994ApJ...421..640N}
\begin{equation}
\label{Tacc}
    T_{\rm acc} = \left[ \left( \frac{9\dot{M}\Omega^2}{32\pi\sigma} \right) \left( \frac{\tau_{\rm mid}}{2} + \frac{1}{\sqrt{3}} \right) \right]^{1/4},
\end{equation}
where $\dot{M}$ is the mass accretion rate, $\sigma$ is the Stefan-Boltzmann constant, and $\tau_{\rm mid} = \kappa_{\rm R}\Sigma_{\rm g}/2$ is the vertical optical depth to the midplane defined by the Rossland mean opacity $\kappa_{\rm R}$.
For simplicity, we adopt a constant $\kappa_{\rm R} = 4.5$ cm$^2$~g$^{-1}$ throughout the disk. 
The adopted opacity value is from the calculation by \citet{1985Icar...64..471P} for $T=150$ K, corresponding to H$_2$O sublimation temperature.
Note that their calculations ignore dust growth and also fix the dust-to-gas surface mass ratio to 0.01 throughout the disk.

\subsection{Dust evolution}
\label{subsec:model_dust}
Our dust evolution model is based on the work of \citet{2012ApJ...752..106O} and \citet{2016A&A...589A..15S}.
This model deals with dust grain size evolution according to the growth of particles that dominate overall mass budget of dust (single-size approximation).
In many cases, the largest particles tend to dominate the dust mass at a particular orbital radius. The dust grain size $a$ in this study represents the maximum size of particles.

The dust evolves through collisional coalescence and fragmentation, inward drifting due to the gas drag, and sublimation/dissociation of nitrogen species.
We calculate the evolution of dust surface density $\Sigma_{\rm d}$ and dust number density $N_{\rm d}$ with the single-size approximation based on the prescription of \citet{2016A&A...589A..15S}. We assume $\Sigma_{\rm d}/\Sigma_{g} = 0.01$ at $t = 0$.
The equation governing $\Sigma_{\rm d}$ is given by
\begin{equation}
\label{Sigma_d_evolution}
     \frac{\partial \Sigma_{\rm d}}{\partial t} +\nabla \cdot \mathcal{M_{\rm d}}= -\sum_i S_{\mathrm{N},i},
\end{equation}
with dust mass flux $\mathcal{M_{\rm d}}$ written as
\begin{equation}
\label{Sigma_d_massflux}
   \mathcal{M_{\rm d}} = r\Sigma_{\rm d}v_{\rm d} - rD_{\rm d}\Sigma_{\rm g} \nabla \left(\frac{\Sigma_{\rm d}}{\Sigma _{\rm g}}\right),
\end{equation}
where $v_{\rm d}$ and $D_{\rm d}$ are the radial velocity and diffusion coefficient of the dust, respectively, and $S_{\mathrm{N},i}$ represents the rate of dust surface density loss due to ammonium salt dissociation  or NH$_3$ ice sublimation (the subscript $i$ stands for either ammonium salt or NH$_3$ ice).
We describe our prescription for $S_{\mathrm{N},i}$ in Section \ref{subsec:model_N_species}.
The dust radial velocity is given by \citep{1972fpp..conf..211W, 1976PThPh..56.1756A, 1977MNRAS.180...57W}
\begin{equation}
\label{v_d}
    v_{\rm d} = -\frac{2\rm{St}}{1+\rm{St}^2} \eta v_{\rm K},
\end{equation}
where St is the Stokes number of the dust particle, $\eta$ is the dimensionless quantity characterizing the deviation of the angular velocity of the gas from the Keplerian, and $v_{\rm K} = r\Omega_{\rm K}$ is the Kepler velocity.
The Stokes number is the product of the Keplerian orbital frequency and the particle's stopping time, which is the timescale of momentum relaxation due to gas drag.
The Stokes number is given by the following equation using the Epstein and Stokes laws:
\begin{equation}
\label{St}
    {\rm St} = \frac{\pi}{2} \frac{\rho_{{\rm s}} a}{\Sigma_{\rm g}} \mathrm{max} \left( 1, \frac{4 a}{9\lambda_{\rm mfp}} \right),
\end{equation}
where $\rho_{\rm s}$ is the dust internal density, $a$ is the dust particle radius corresponding to the dust mass $m_{\rm d} = \Sigma_d / N_{\rm d}$, and $\lambda_{\rm mfp} = \mu m_{\rm H} / \sigma_{\rm mol} \rho_{\rm g}$ is the mean free path of a gas particle with the collisional cross-section $\sigma_{\rm mol}$ (taken to be $2.0 \times 10^{-15}~{\rm cm}^2$) and gas density $\rho_{\rm g} = \Sigma_{\rm g}/\sqrt{2\pi}H_{\rm g}$. Here, $H_{\rm g} = c_{\rm s}/\Omega_{\rm K}$ is the gas scale height. The dimensionless parameter $\eta$ is given by
\begin{equation}
\label{eta}
    \eta = -\frac{1}{2}\left(\frac{c_{\rm s}}{v_{\rm K}}\right)^2 \frac{{\rm d} \ln{(c_{\rm s}^2 \rho_{\rm g})}}{\rm{d} \ln r}.
\end{equation}
The dust diffusion coefficient is given by
\begin{equation}
    D_{\rm d} = \frac{D_{\rm g}}{1+\rm{St}^2},
\end{equation}
where $D_{\rm g}$ is the gas diffusion coefficient and we assume $D_{\rm g} = \nu_{\rm visc}$.
Equation \eqref{v_d} does not include the term due to the gas radial velocity $v_{\rm g}$ \citep{2002ApJ...581.1344T} and is justified when radial drift is predominant
\footnote{From equation \eqref{v_g}, $v_{\rm g} \sim \nu_{\rm visc}/r = \alpha c_{\rm s}^2/v_{\rm K}$, and from equations \eqref{v_d} and \eqref{eta}, $v_d \sim {\rm St} c_{\rm s}^2/v_{\rm K}$ when ${\rm St} < 1$. 
Therefore, whether the gas-induced or drift-induced term in the dust velocity is dominant depends on the value of $\alpha$ and St.
In our model, dust growth causes St to increase immediately and the drift becomes predominant. \label{vd_justification}}.

The dust number density $N_{\rm d}$ is evolved by the following advection-diffusion equation including dust coalescence and fragmentation
\begin{equation}
\label{N_d_evolution}
     \frac{\partial N_{\rm d}}{\partial t} +\nabla \cdot \mathcal{N_{\rm d}}= \frac{\Delta N_{\rm d}}{t_{\rm{coll}}},
\end{equation}
with 
\begin{equation}
\label{N_d_numberflux}
    \mathcal{N_{\rm d}} = rN_{\rm d}v_{\rm d} - rD_{\rm d}\Sigma_{\rm g} \nabla \left(\frac{N_{\rm d}}{\Sigma _{\rm g}}\right),
\end{equation}
where $\Delta N_{\rm d}$ is the dust number density change in a collision and $t_{\rm coll}$ is the mean collision time.
We model $\Delta N_{\rm d}$ as \citep{2009ApJ...702.1490W,2012ApJ...753L...8O, 2016ApJ...821...82O}
\begin{equation}
\label{Delta_N}
    \Delta N_{\rm d} = \min \left[1, \frac{\ln(\Delta v/\Delta v_{\rm{frag}})}{\ln 5}\right]N_{\rm d},
\end{equation}
where $\Delta v_{\rm frag}$ and $\Delta v$ are the fragmentation velocity and relative velocity of dust particle, respectively.
Equation \eqref{Delta_N} determines whether fragmentation or coalescence occurs depending on the magnitude of $\Delta v$ and $v_{\rm frag}$: fragmentation ($\Delta N > 0$) when $\Delta v > v_{\rm frag}$, coalescence ($\Delta N < 0$) when $\Delta v < v_{\rm frag}$.
The relative velocity $\Delta v$ includes five components induced by Brownian motion $\Delta v_{\rm B}$, radial drift $\Delta v_{\rm d}$, azimuthal drift $\Delta v_{\phi}$, vertical settling $\Delta v_{\rm z}$, and turbulence $\Delta v_{\rm t}$, and given by
\begin{equation}
\label{Delta_v}
    \Delta v = \sqrt{(\Delta v_{\rm B})^2 + (\Delta v_{\rm d})^2 +(\Delta v_{\rm \phi})^2 +(\Delta v_{\rm z})^2 + (\Delta v_{\rm t})^2}.
\end{equation}
We evaluate each velocity components with the equations given in \citet{2012ApJ...752..106O}.
To calculate the relative velocities, we need the Stokes numbers ${\rm St}_1$ and ${\rm St}_2$ for the colliding particles. 
Following \citet{2016A&A...589A..15S}, we take $\mathrm{St}_1 = \mathrm{St}(r)$ and $\mathrm{St}_2 = 0.5 \mathrm{St}(r)$ to take into account the finite width of the dust grain size distribution, which is not resolved in our single-size approximation.

The mean collision time is given by
\begin{equation}
\label{t_coll}
    t_{\rm coll} = \frac{H_{\rm d}}{2 \pi a^2 \Delta v N_{\rm d}},
\end{equation}
where $H_{\rm d}$ is the dust scale height written as \citep{1995Icar..114..237D, 2007Icar..192..588Y}
\begin{equation}
\label{H_d}
    H_{\rm d} = H_{\rm g} \left(1+ \frac{\rm St}{\alpha}\frac{1+2\rm{St}}{1+\rm{St}}\right)^{-1/2}.
\end{equation}

By solving equations \eqref{Sigma_d_evolution} and \eqref{N_d_evolution}, we calculate the dust surface density and dust size using $m_{\rm d} = \Sigma_{\rm d}/N_{\rm d}$ and $m_{\rm d} = 4 \pi a^3 \rho_{\rm s}/3$.

\subsection{Production and transport of nitrogen-bearing vapor}
\label{subsec:model_N_species}
We define $N_{{\rm NH}_3,{\rm vap}}$ and $N_{{\rm NH_4X},{\rm vap}}$ as the number surface densities of NH$_3$ vapor produced by NH$_3$ ice sublimation and ammonium salt (NH$_4$X) dissociation.
In this paper, the term "NH$_3$ ice line" and "ammonium salt line" refer to the orbitals of NH$_3$ sublimation and ammonium salt dissociation, respectively.
The evolution of $N_{i,{\rm vap}}$ ($i=$ NH$_3$ and NH$_4$X) obeys an advection--diffusion equation with a nitrogen source term analogous to Equation \eqref{Sigma_d_evolution},
\begin{equation}
\label{Sigma_i,vap_evolution}
     \frac{\partial N_{i, {\rm vap}}}{\partial t} +\nabla \cdot \mathcal{N}_{i,{\rm vap}}= \sum_i \left(\frac{S_{\mathrm{N}, i}}{m_i}\right),
\end{equation}
with
\begin{equation}
\label{Sigma_i,vap_massflux}
   \mathcal{N}_{i,{\rm vap}} = rN_{i,{\rm vap}}v_{\rm g} - rD_{\rm d}\Sigma_{\rm g} \nabla \left(\frac{N_{i, {\rm vap}}}{\Sigma _{\rm g}}\right),
\end{equation}
where $m_{\rm i}$ is the molecular mass of the nigrogen-bearing species $i$.

In this study, we adopt the following simplified prescription for $S_{\mathrm{N},i}$,
\begin{equation}
\label{S_N}
    S_{\mathrm{N},i} = 
    \left\{
   \begin{array}{ll}
       {\displaystyle f_{i,\rm{dust}}}{\displaystyle\frac{\mathcal{M}_{\rm{d}}}{r\Delta r}}, & T = T_{\mathrm{vap},i}, \\
       0,        & \rm{otherwise},
   \end{array}
   \right.
\end{equation}
where $f_{i,\rm{dust}}$, and $T_{\mathrm{vap},i}$ are the mass fraction in the dust and dissociation/sublimation temperature of the nigrogen-bearing species $i$, respectively, and $\Delta r$ is the cell width.
This formulation implies that solid nitrogen molecules contained in dust are instantaneously converted into vapor upon crossing the NH$_3$ ice line or the ammonium salt line.
In general, similar models that describe the evolution of dust composition determine sublimation and recondensation rates based on the saturation vapor pressure of each chemical species \citep{2017A&A...602A..21S,2019MNRAS.487.3998B, 2021A&A...654A..71S}. Our model corresponds to a case where the recondensation rate is zero at the snow and dissociation lines, while the sublimation rate is sufficiently high for the chemical species to be immediately lost from the dust.
Vapor release occurs only at the NH$_3$ ice line or at the ammonium salt line. Furthermore, we assume that NH$_3$ and ammonium salts are absent within these orbits under initial conditions.
NH$_3$ vapor can be converted to highly volatile chemical species such as atomic N and N$_2$ by photodissociation \citep{2014ApJ...797..113S}.
The disk observations of \citet{2019ApJ...874...92P} support that NH$_3$ depletion in the inner region of the disk ($r < 10$ au) comes from the destruction of NH$_3$ vapor by photodissociation.
We neglect recondensation of the produced NH$_3$ vapor outside the NH$_3$ ice line in a fiducial model.

On the other hand, if photodissociation cannot efficiently destroy NH$_3$ vapor, recondensation of NH$_3$ vapor outside the NH$_3$ ice line may be important for nitrogen enrichment in the disk/planet.
For comparison with the fiducial model, we also perform a model calculation considering the recondensation of NH$_3$.
In this case, we adopt the following mass flux for the vapor produced in the NH$_3$ ice line and the ammonium salt line, modified from equation \eqref{Sigma_i,vap_massflux},
\begin{equation}
\label{Sigma_i,vap_massflux_recondensation}
   \mathcal{N}_{i,{\rm vap}} = rN_{i,{\rm vap}}v_{\rm rec} - rD_{\rm d}\Sigma_{\rm g} \nabla \left(\frac{N_{i, {\rm vap}}}{\Sigma _{\rm g}}\right),
\end{equation}
with
\begin{equation}
\label{v_rec}
    v_{\rm rec} = 
    \left\{
   \begin{array}{ll}
       v_{\rm g}, & T \ge T_{\rm vap,NH_3}, \\
       v_{\rm d}, & T < T_{\rm vap,NH_3}.
   \end{array}   
   \right.
\end{equation}
Using this equation, the NH$_3$ vapor has the dust velocity $v_{\rm d}$ outside the NH$_3$ ice line and drifts inward (i.e., we reproduce the freezing of the vapor onto the dust).
Note that in this model, $\Sigma_{i, {\rm vap}}$ includes NH$_3$ ice that has recondensed outside the NH$_3$ ice line.

To determine the nitrogen content of the dust, we assume that the dust grains are composed of ammonium salts, refractories (organic matter and minerals), and volatile ices.
For the sake of convenience, we collectively refer to salts and refractories as ``rock''.
The rock-to-ice mass ratio of the grains is assumed to be 4, consistent with the observed value in the coma of comet 67P \citep{2015Sci...347a3905R}. 
Assuming the salt-to-rock mass ratio of 0.1--0.3 estimated for the 67P by \citet{2020Sci...367.7462P}, the mass fraction of the salts in the grains can be estimated to be 0.08--0.24.
Consequently, we set $f_{\rm NH_{4}X,dust} = 0.1$, 0.2, and 0.3. 
In a fiducial model, all salts are assumed to be ammonium formate (NH$_4$X = NH$_4$HCO$_2$) with the dissociation temperature of 200 K \citep{2016ApJ...829...85B}.
To explore the dependence of our results on the dissociation temperature, we also consider NH$_4$CN with a lower dissociation temperature of 150 K \citep{2011A&A...535A..47D}.
For NH$_3$ ice, we set $f_{\rm NH_{3},dust}$ = 0.005 and assume the NH$_3$ sublimation temperature of 80 K \citep{2016ApJ...821...82O}.
Because of the large rock-to-ice mass ratio, we set the dust internal density to $\rho_{\rm s} = 3~\rm g~cm^{-3}$. 
We do not account for nitrogen delivered by the refractory dust component.

Our adopted dust composition based on the dehydrated grains of 67P may overestimate the salt-to-ice mass ratio relative to the actual dust composition in disks.
Assuming that the primary ice component is H$_2$O and the salt is NH$_4$HCOO, then NH$_4^+$/H$_2$O $\sim$ 0.3 when $f_{\rm NH_{4}X,dust} = 0.2$. 
In comparison, interstellar ices have lower nitrogen-to-H$_2$O ratios, with NH$_4$/H$_2$O $\sim$ 0.1 and (NH$_4^+$ + NH$_3$)/H$_2$O $\sim$ 0.2 \citep{2023NatAs...7..431M}.
Therefore, our model implicitly assumes that the grains have lost H$_2$O during disk formation or radial transport to obtain a 67P-like composition.
The variation in the salt-to-ice ratio modulates the oxygen delivery to Jupiter, constituting a pivotal factor regarding Jupiter's oxygen enrichment.
The impact of varying dust-to-ice mass fractions is discussed in Section \ref{subsec:discussion_model}.

\subsection{Planetary nitrogen abundance}
\label{subsec:model_planet}
We calculate the planetary nitrogen abundance sequentially from the disk evolution.
For computational simplicity, we insert a protoplanet at time $t = t_{\rm p}$ and orbital radius $r = r_{\rm p}$, reset the inner boundary of the gas disk to the core's orbit, and let the core accrete all the disk gas flowing to the inner computational boundary by applying the zero-torque boundary condition.
The zero-torque boundary condition of the gas produces a pressure bump outside the inner boundary, which causes dust pile-up. This effect reproduces gap formation.
We neglect core growth, planetary migration, and gas inflow from the inner region of the core's orbit.
This treatment is based on the runaway accretion scheme of \citet{2016ApJ...823...48T, 2020ApJ...891..143T}.
They show analytically that a planet core with a 30 $M_{\oplus}$--$10M_{\rm jup}$ would capture almost all gas inflowing into the planetary gap, where $M_{\rm jup}$ represents Jupiter's mass.
In our calculations, the protoplanet is assumed to have the initial mass $M_{\rm p,ini}$ of 30 $M_{\oplus}$ and comprises a heavy element core and an envelope.
The mass of the heavy-element core is taken to be 13 $M_{\oplus}$ (0.04 $M_{\rm jup}$) based on Juno's constraint that sets an upper limit of 6--27 $M_{\oplus}$ for the total heavy-element mass of Jupiter \citep{2017GeoRL..44.4649W,2019A&A...632A..76N}.
However, Jupiter's internal structure model based on the Juno's gravity data suggests that Jupiter's core is diluted
\citep[e.g.,][]{2018A&A...610L..14V, 2019ApJ...872..100D, 2022Icar..37814937H}.
Furthermore, Jupiter's formation models show that once the core grows into a few $M_{\oplus}$ in mass and acquires an envelop, heavy elements tend to dissolve in the envelope, irrespective of whether the solids are brought about by pebble accretion or planetesimal accretion \citep{2007Icar..187..600I, 2011MNRAS.416.1419H, 2018A&A...611A..65B, 2020ApJ...900..133V}.
Therefore, we assume that 10 $M_{\oplus}$ out of 13 $M_{\oplus}$ of heavy elements in the core dissolve into the atmosphere.
Several long-term evolutionary models of Jupiter's interior structure have demonstrated that convective mixing erases compositional gradients in the outer regions of the planet (50\% or more of the radius) on scales of millions of years \citep{2018A&A...610L..14V, 2020A&A...638A.121M}. We assume that heavy elements dissolved in the atmosphere undergo homogenization through convective mixing.
On the other hand, we note that the composition of actual Jupiter is not entirely uniform in the vertical direction, as evidenced by Jupiter having a diluted core. A discussion of compositional heterogeneity is provided in Section \ref{subsec:discussion_model}.

We continue calculating the time evolution of dust and NH$_3$ vapor using equations \eqref{Sigma_d_evolution}, \eqref{N_d_evolution} and \eqref{Sigma_i,vap_evolution}, but with the inner boundary condition at $r=r_{\rm p}$.
If the ammonium salt line is located inside the planetary core orbit, salt dissociation does not contribute to NH$_3$ vapor abundance after core formation.
For the dust flux at the inner boundary, we adopt a zero-flux condition because of the lack of background gas inside the boundary, i.e., no radial drift.
The inner boundary flux of NH$_3$ vapor is determined solely from the advection term: the first term on the right-hand side of equation \eqref{Sigma_i,vap_massflux}.

The accreting gas mass on the planet, $M_{\rm p, atm}(t)$, is
\begin{equation}
\label{Mp,atm}
    M_{\rm p, atm}(t) = \int_{t_{\rm p}}^{t} 2\pi r_{\rm p} \left[(v_{\rm g}\Sigma_{\rm g})_{r = r_{\rm p}} + \sum\limits_i (v_{\rm g}\Sigma_{i, {\rm vap}})_{r = r_{\rm p}}\right] dt'.
\end{equation}
The number of NH$_3$ molecules accreting on the planet, $n_{\rm NH_3, atm}$, is
\begin{equation}
\label{MN,atm}
    n_{\rm NH_3, atm}(t) = \int_{t_{\rm p}}^{t} 2\pi r_{\rm p} \left[\sum\limits_i f_{\mathrm{N}, i}(v_{\rm g}N_{i, {\rm vap}})_{r = r_{\rm p}} \right] dt',
\end{equation}
where, $f_{\mathrm{NH_3}, i}$ is a mass fraction of NH$_3$ in the molecule (For NH$_4$HCO$_2$, $f_{\rm NH_3, NH_4HCO_2} = 0.3$).
Thus, the planetary atmosphere's nitrogen atomic abundance is 
\begin{equation}
x_{\rm N,p}(t) = \frac{(n_{\rm NH_3, atm} + n_{\rm NH_3,core})}{(n_{\rm H, atm} + 3n_{\rm NH_3, atm} + 3n_{\rm NH_3,core})},
\end{equation}
where $n_{\rm NH_3,core} = 10f_{\rm NH_{4}X,dust}f_{\rm NH_3, NH_4X}M_{\oplus}/(17m_{\rm H})$ is the number of NH$_3$ molecules dissolved from the core to the atmosphere and $n_{\rm H, atm} = M_{\rm p, atm} / m_{\rm H}$ is the number of hydrogen in the atmosphere derived from the back ground gas.
During the gas accretion phase, the planet exceeds the pebble isolation mass \citep{2014A&A...572A..35L}, so we neglect the core growth by solid accretion.
We also assume that the protoplanet forms at 0.6 Myr at $r_{\rm p}= 3.0$ au.
This planet's orbit is taken to be near the water-snow line (150--170 K), which is favorable for planetesimal formation \citep{2017A&A...608A..92D}. Moreover, this orbit is near the ice crystalline-amorphous phase transition line (143 K) and is favorable for the hypervolatile element enrichment proposed by \citet{2019ApJ...875....9M}.

\subsection{Numerical settings and parameter set}
\label{subsec:model_settings}
We numerically solve equations \eqref{Sigma_g_evolution}, \eqref{Sigma_d_evolution}, \eqref{N_d_evolution}, and \eqref{Sigma_i,vap_evolution} with the finite volume method. The computational domain ranges from 0.5 to 200 au and is divided into 800 radial cells with an equal logarithmic length.

The simulation consists of several models with different parameter sets.
We set up as a fiducial model when a core forms at 0.6 Myr at 3.0 au in a disk with moderate turbulence strength $\alpha=10^{-3}$, a fragmentation velocity of icy dust $v_{\rm frag}=10$ m s$^{-1}$, and salt dissociation temperature $T_{\rm vap, NH_4X} = 200$ corresponding to ${\rm X} = {\rm HCO_2}$.
Other parameter sets are shown in Table \ref{tb:parameter}, with seven cases in which the protoplanet formation age, core orbit, salt dissociation temperature, turbulence strength, or dust fragmentation velocity differ from the fiducial model.
In addition to these parameter sets, we calculate Case 8, which considers the effect of recondensation according to equations \eqref{Sigma_i,vap_massflux_recondensation} and \eqref{v_rec}.
For each case, we take the ammonium salt content in the dust to be 10, 20, or 30 wt\% (corresponding to $f_{\rm NH_{4}X,dust} = 0.2^{+0.1}_{-0.1}$).
In all cases, the simulations run until the planet reaches Jupiter's mass.

\begin{table*}[t]
    \caption{Parameter Sets Adopted in the Simulations}
    \begin{tabular}{cccccc}
     Case & $t_{\rm p}$ (Myr) & $\alpha$ & $r_{\rm p}$ (au) & $v_{\rm frag}$ (m s$^{-1}$) & $T_{\rm vap, NH_4X}$ (K) \\ \hline
     Fiducial & 0.6 & 10$^{-3}$ & 3.0 & 10.0  & 200\\
     Case~1 & \textbf{0.1} & 10$^{-3}$ & 3.0 & 10.0 & 200\\
     Case~2 & \textbf{1.0} & 10$^{-3}$ & 3.0 & 10.0 & 200 \\
     Case~3 & 0.6 & 10$^{-3}$ & \textbf{5.2} & 10.0 & 200\\ 
     Case~4 & 0.6 & 10$^{-3}$ & \textbf{10.0} & 10.0 & 200\\
     Case~5 & 0.6 & 10$^{-3}$ & 3.0 & 10.0 & \textbf{150} \\
     Case~6 & 0.6 & 10$^{-3}$ & 3.0 & \textbf{1.0}  & 200\\  
     Case~7 & 0.6 & \textbf{10$^{-4}$} & 3.0 & 10.0 & 200\\
     Case~8 & \multicolumn{5}{c}{Same as Fiducial, but equation \eqref{Sigma_i,vap_massflux_recondensation} is adopted for $\mathcal{N}_{i,{\rm vap}}$} \\ \hline
    \end{tabular}
    \label{tb:parameter}
\end{table*}

\section{Results}
\label{sec:results}
Here we show the results of our simulations.
In Section \ref{subsec:result_disk}, we use the results of the fiducial case to see the production and diffusion of nitrogen vapor from dust and the time evolution of the nitrogen abundance of planets forming in the inner region of the disk (r = 3 au).
In Section \ref{subsec:parameter_survey}, we carry out a parameter study for the case listed in Table 1 to investigate the effect of the model parameters on the availability of nitrogen vapor in the disk and on the abundance of planetary nitrogen.

\subsection{The fiducial model}
\label{subsec:result_disk}

\subsubsection{Time evolution of the disk until the core formation}
\label{subsubsec:result_disk_before}

\begin{figure}[t]
\includegraphics[width=\hsize, bb=0 0 532 382]{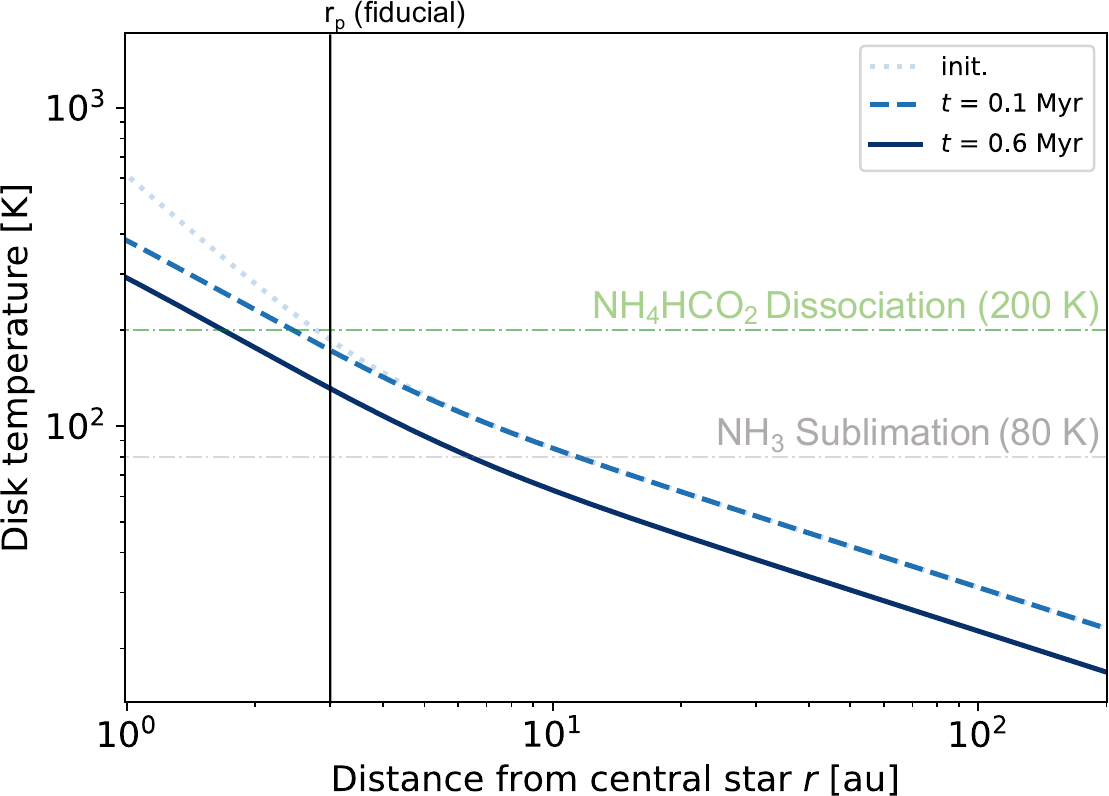}
\caption{Time evolution of the disk temperature $T$ from the fiducial model as a function of the distance from the central star $r$. The blue dotted, dashed, and solid lines are the snapshots at times $t = 0, 0.1$ and 0.6 Myr, respectively. The gray and green dash-dotted lines represent the sublimation temperature of NH$_3$ and the dissociation temperature of NH$_4$HCO$_2$, respectively. {Alt text: Line graph showing the disk temperature evolution.}}
\label{fig:time_evo_T}
\end{figure}

We begin by describing disk evolution before the insertion of the gas giant core.
Figure \ref{fig:time_evo_T} shows the radial distribution of the disk temperature $T$ at different times from the fiducial model.
The slope of the disk temperature depends on whether irradiation or viscous heating is dominant.
In this study, viscous heating dominates at $r \lesssim$ 3 au.
Because the fiducial model takes the dissociation temperature of ammonium salts to be 200 K, viscous heating determines the location of the dissociation line.
Due to the temperature evolution associated with decreases in the mass accretion rate of the gas, the salt dissociation line migrates inward from 2.8 to 1.7 au in the first 0.6 Myr of disk evolution.
The NH$_3$ sublimation line is located in the irradiation-dominated region, moving inward from 11.6 to 6.4 au.

\begin{figure*}[t]
    \begin{tabular}{cc}
    \hspace{-0.5cm}
      \begin{minipage}[t]{0.45\linewidth}
        \centering
        \includegraphics[width=\linewidth, bb=0 0 455 305]{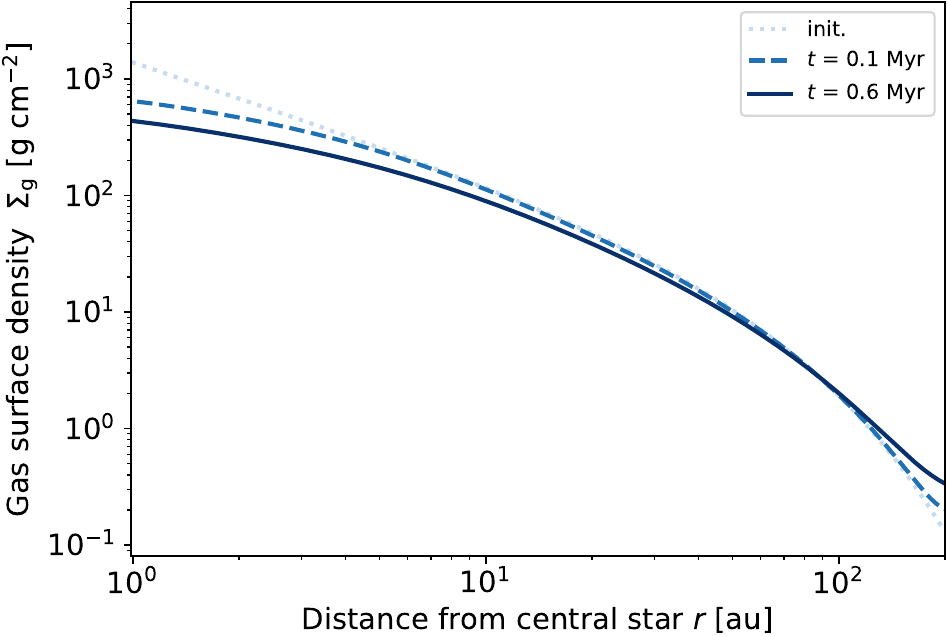}
      \end{minipage} &
        \hspace{0.5cm}
      \begin{minipage}[t]{0.45\linewidth}
        \centering
        \includegraphics[width=\linewidth, bb=0 0 453 305]{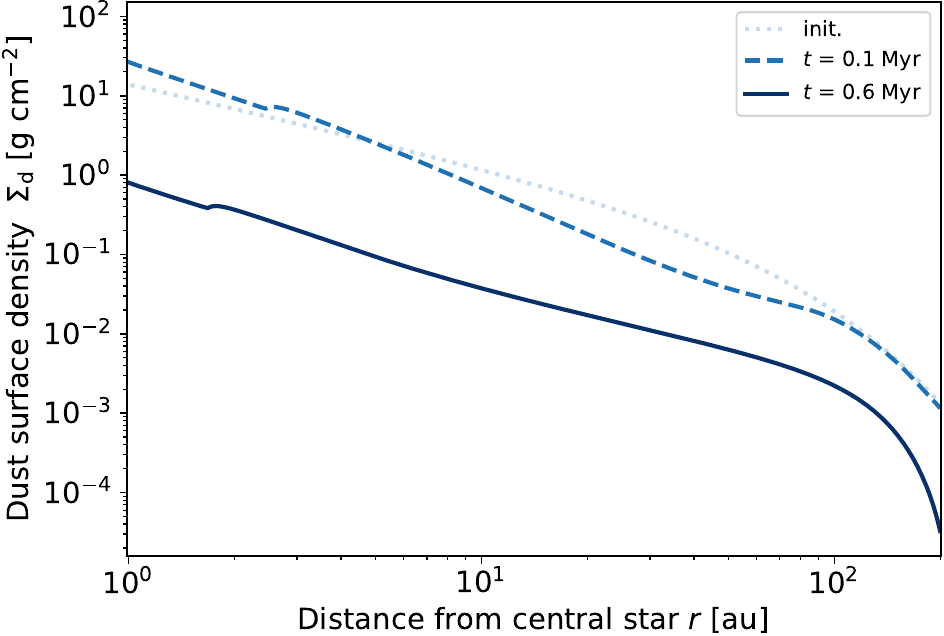}
      \end{minipage} \\
      \hspace{-0.5cm}
      \begin{minipage}[t]{0.45\linewidth}
        \centering
        \includegraphics[width=\linewidth, bb=0 0 491 333]{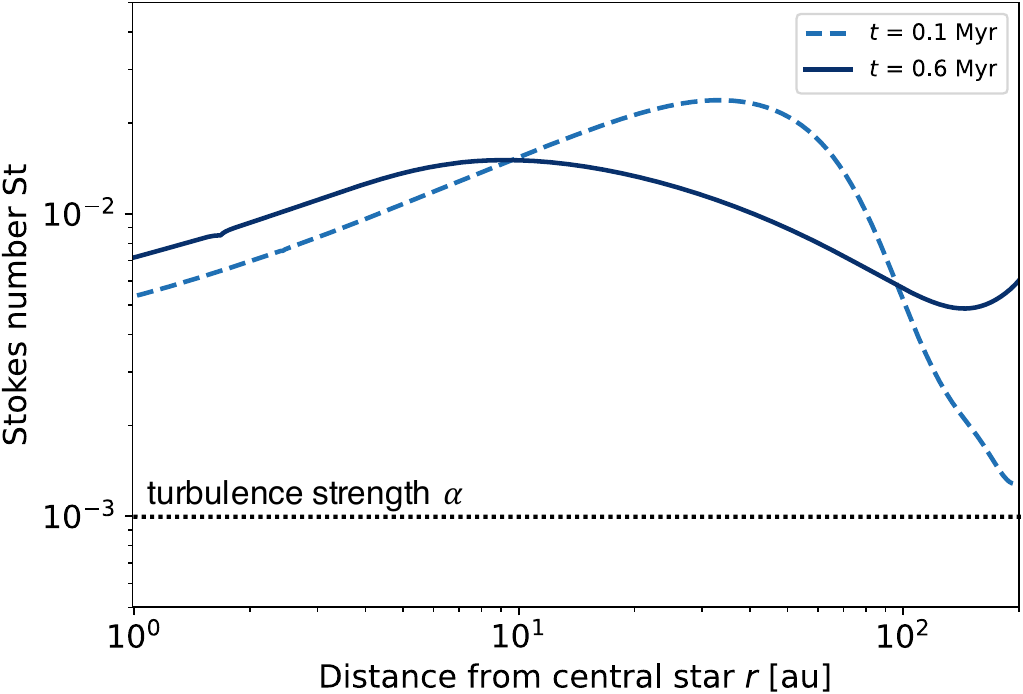}
      \end{minipage} &
        \hspace{0.5cm}
      \begin{minipage}[t]{0.45\linewidth}
        \centering
        \includegraphics[width=\linewidth, bb=0 0 450 307]{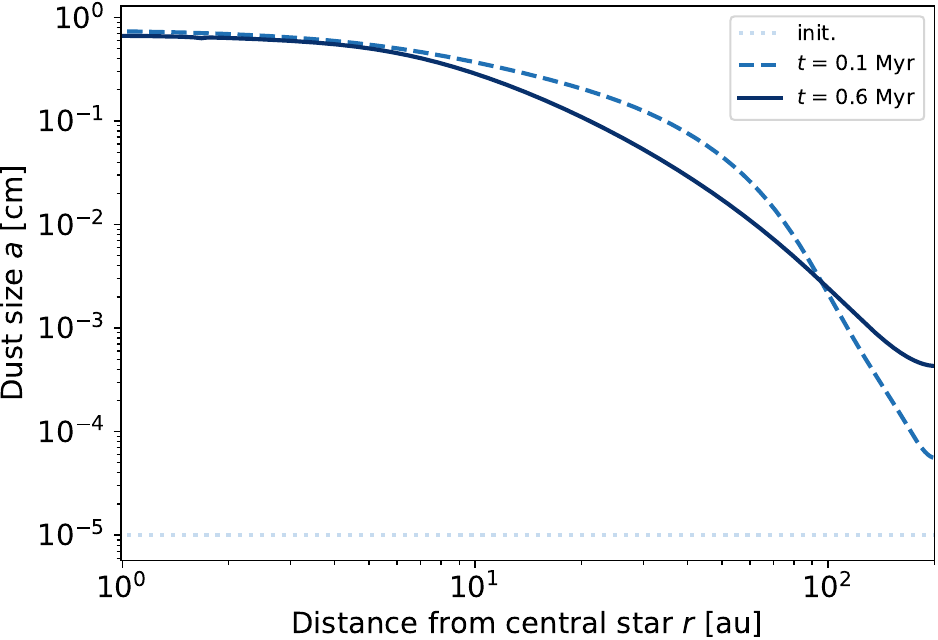}
      \end{minipage} 
    \end{tabular}
    \caption{Time evolution of the gas surface density $\Sigma_{\rm g}$ (upper left), the dust surface density $\Sigma_{\rm d}$ (upper right), the Stokes number St (lower left), and the dust grain size (lower right), respectively. The results are from the fiducial model. The blue dotted, dashed, and solid lines are the snapshots at times $t = 0, 0.1$ and 0.6 Myr, respectively. The small drops in $\Sigma_{\rm d}$ at 2.5 au (on the dashed line) and 1.8 au (on the solid line) are caused by ammonium salt dissociation. {Alt text: Four line graphs showing the evolution of physical properties of gas and dust.}}
\label{fig:time_evo_disk}
\end{figure*}

The upper panels of Figure \ref{fig:time_evo_disk} show the time evolution of the gas and dust surface densities $\Sigma_{\rm g}$ and $\Sigma_{\rm d}$ in the fiducial model.
Due to viscous diffusion, the gas surface density inside and outside the initial characteristic radius $r_{\rm c} = 50~ \rm au$ decreases and increases over time, respectively.
The dust surface density decreases at a higher rate owing to the grains' radial inward drift.
The shallow drop of $\Sigma_{\rm d}$ at 2 au is caused by the loss of the dust mass due to salt dissociation.

The lower panels of Figure \ref{fig:time_evo_disk} show the time evolution of the Stokes number St and size $a$ of the dust grains.
Dust grains grow until  their mutual collision velocity $\Delta v$ reaches the fragmentation threshold $v_{\rm frag}$.
The grains interior to 20 au reach the fragmentation-limited size within 0.1 Myr.
Salt dissociation and NH$_3$ ice sublimation have little effect on $a$.
As the dust grains grow, their Stokes number exceeds 10$^{-3}$ throughout the disk at 0.1 Myr. The maximum Stokes number satisfies the requirement ${\rm St} > \alpha$ for the dust velocity to be dominated by radial drift (see Sect \ref{subsec:model_dust} and footnote \ref{vd_justification}), in particular at $r < 10$ au.

\begin{figure}[t]
\includegraphics[width=\hsize,bb=0 0 532 383]{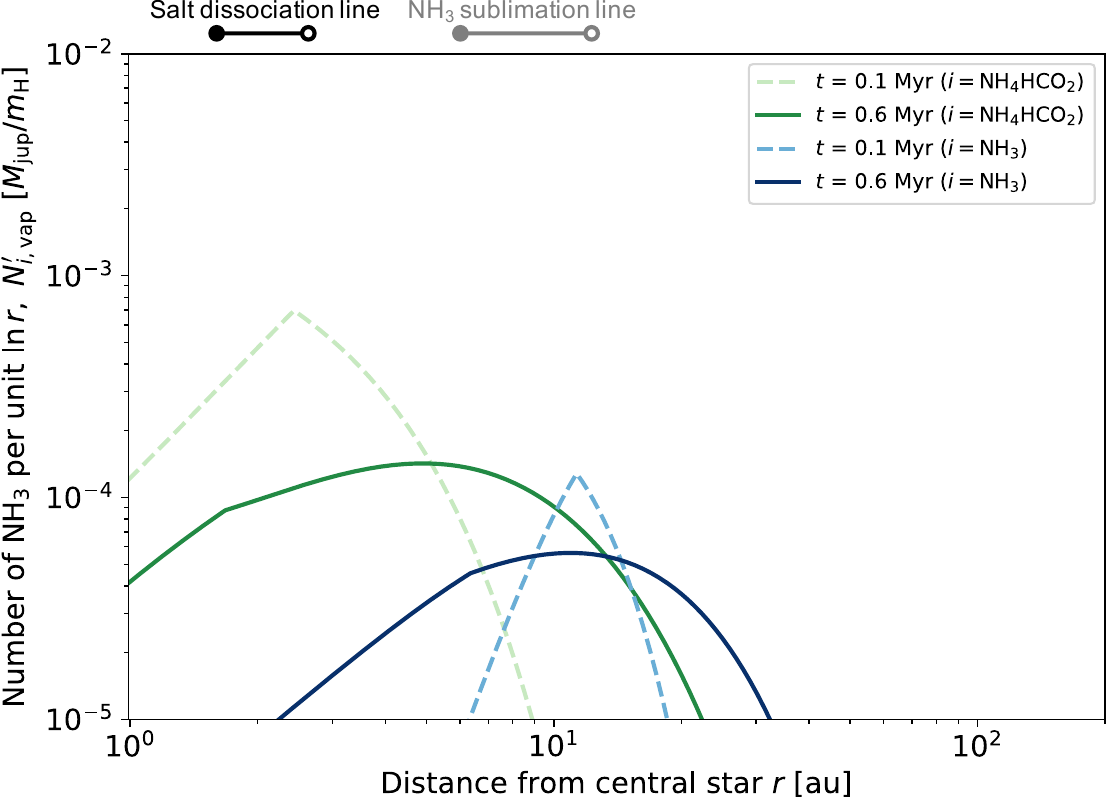}
\caption{Time evolution of the number of NH$_3$ per unit $\ln r$ derived from NH$_4$HCO$_2$ salt $N_{\rm NH_4HCO_2, vap}'$ (green) and NH$_3$ ice $N_{\rm NH_3, vap}'$ (blue) from the fiducial model. The dashed and solid lines are the snapshots at $t = 0.1$ and 0.6 Myr.
The open and filled circles above the panel indicate the locations of the NH$_3$ ice line (gray) and the ammonium salt line (black) at $t = 0$ and $t = 0.6$ Myr. {Alt text: Line graph showing the distribution of NH$_3$ molecules produced by dissociation or sublimation.}}
\label{fig:time_evo_r2Sigma}
\end{figure}

To see how much nitrogen vapor is produced from the dust and how the vapor diffuses over the disk, we plot in Figure \ref{fig:time_evo_r2Sigma} the number of NH$_3$ molecules per unit $\ln r$, $N_{i, \rm vap}'(r) \equiv 2\pi r^2 N_{i,{\rm vap}}$, from the fiducial model with $f_{\rm NH_4X,dust} = 0.2$.
At $t < 0.1$ Myr, $N_{\rm NH_4HCO_2, vap}'$ and $N_{\rm NH_3, vap}'$ have peaks at 2.5 and 11 au, which originate from salt dissociation and NH$_3$ ice sublimation, respectively.
The peak value at the salt line is one order of magnitude higher than that at the NH$_3$ sublimation line. This reflects the amount of   NH$_3$ in the salt, $f_{\rm NH_4X,dust} \times f_{\rm{NH_3, NH_4HCO_2}} = 0.06$, versus that of NH$_3$ ice in the dust, $f_{\rm NH_3, dust} = 0.005$.
These peaks are smeared out on the diffusion timescale $\sim r^2/D_{\rm g} \sim 0.25$ Myr around the salt line.
As time passes, the peak of $N_{\rm NH_4HCO_2, vap}'$ moves outward because of turbulent diffusion.

\subsubsection{Gas accretion onto the protoplanet}
\label{subsubsec:result_planet}

\begin{figure*}[t]
    \begin{tabular}{cc}
    \hspace{-0.7cm}
      \begin{minipage}[t]{0.475\linewidth}
        \centering
        \includegraphics[width=\linewidth, bb= 0 0 447 305]{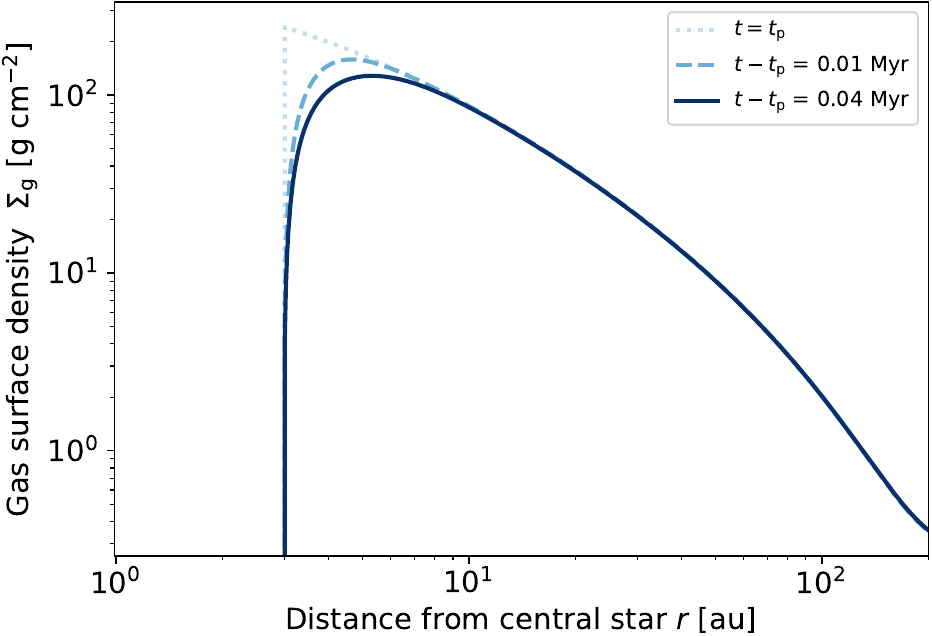}
      \end{minipage} &
        \hspace{0.5cm}
      \begin{minipage}[t]{0.475\linewidth}
        \centering
        \includegraphics[width=\linewidth, bb= 0 0 531 365]{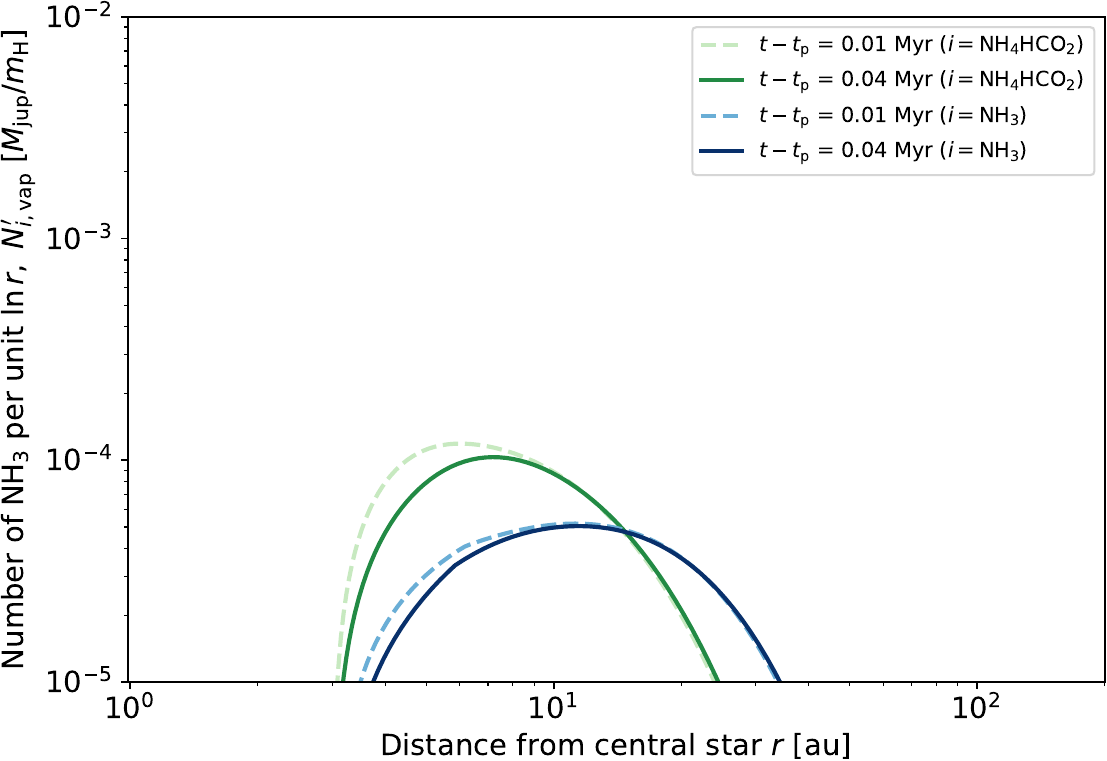}
      \end{minipage}
    \end{tabular}
    \caption{Time evolution of the gas surface density $\Sigma_{\rm g}$ (left) and $N_{i, \rm vap}'$ (right) after the core insertion. The core is inserted at $t_{\rm p} =$ 0.6 Myr and $r_{\rm p} = $ 3.0 au. These results are from the fiducial model.
    The dotted, dashed, and solid lines are the snapshots at times $t-t_{\rm p} = 0, 0.01$ and 0.04 Myr ($M_{\rm p} = M_{\rm jup}$), respectively. Here, $M_{\rm p}$ is the total planetary mass as $M_{\rm p} = M_{\rm p,atm} + M_{\rm p,core}$. {Alt text: Two line graphs showing the time evolution of the gas surface density and the number of NH$_3$ molecules after the core insertion.}}
\label{fig:time_evo_disk_after}
\end{figure*}

\begin{figure*}[t]
    \begin{tabular}{cc}
    \hspace{-0.7cm}
      \begin{minipage}[t]{0.475\linewidth}
        \centering
        \includegraphics[width=\linewidth, bb= 0 0 489 336]{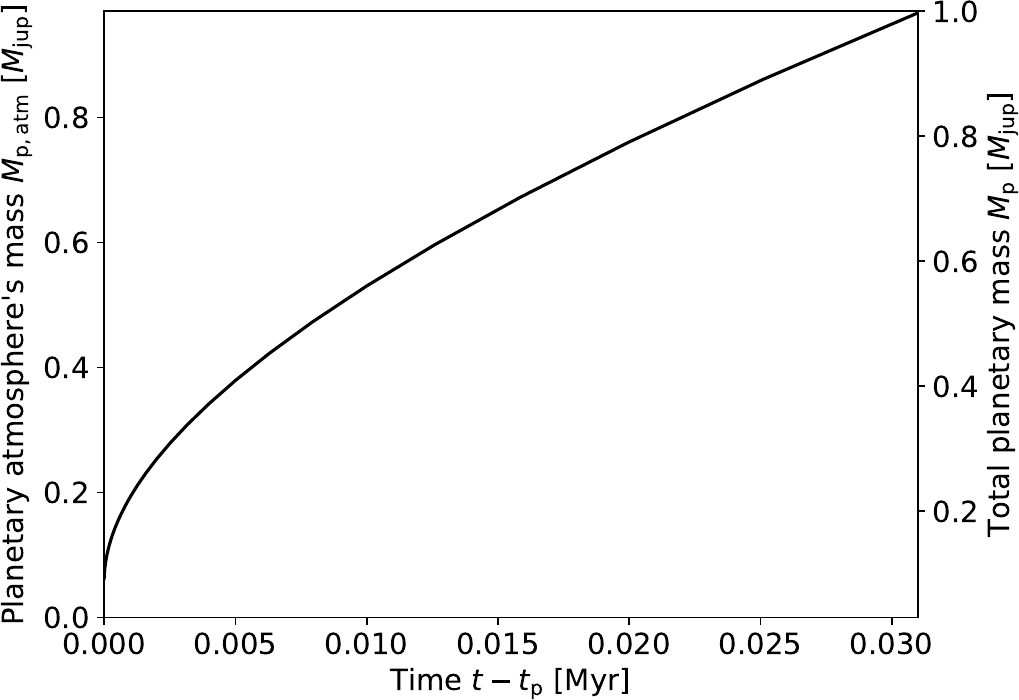}
      \end{minipage} &
        \hspace{0.5cm}
      \begin{minipage}[t]{0.475\linewidth}
        \centering
        \includegraphics[width=\linewidth, bb= 0 0 540 426]{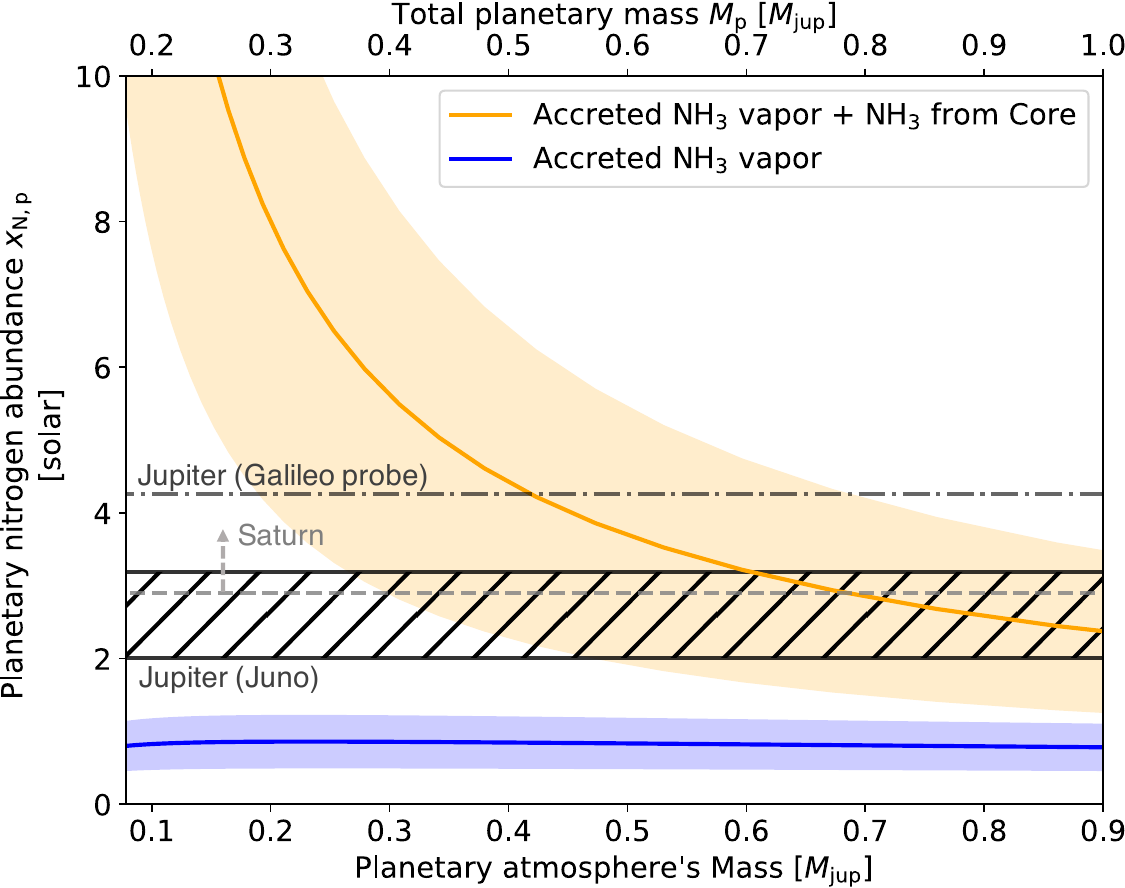}
      \end{minipage}
    \end{tabular}
    \caption{Mass of the planet as a function of time after core insertion $t-t_{\rm p}$ (left) and atmospheric nitrogen abundance of the planet $x_{\rm{N,p}}$ normalized by solar value of $7.78\times10^{-5}$ \citep{2021A&A...653A.141A} as a function of $M_{\rm p}$ (right) from the fiducial model. 
    The orange line in the right panel shows the planetary nitrogen abundance $x_{\rm N,p}$, with the orange shaded area representing the 10 wt\% uncertainty in the NH$_4$HCO$_2$ salt content in comets \citep{2020Sci...367.7462P}. The blue line indicates the contribution of accreted NH$_3$ vapor only ($n_{\rm NH_3,atm} / (n_{\rm H,atm}+3n_{\rm NH_3,atm})$). 
    The black hatched bar marks Jupiter's atmospheric nitrogen abundance including observational errors from Juno \citep{2017Sci...356..821B}. The dot-dashed line marks Jupiter's nitrogen abundance from Galileo probe \citep{2004Icar..171..153W}; this value is highly uncertain ($4.2^{+1.6}_{-1.6}$), but the uncertainty is not indicated in the figure for clarity. The light gray line marks the lower limit of Saturn's nitrogen abundance from Cassini \citep{2009Icar..202..543F}. {Alt text: Two line graphs showing the evolution of planetary mass and atmospheric nitrogen abundance.}}
\label{fig:planetary_growth}
\end{figure*}

Here we describe disk evolution after core insertion and gas accretion by the core. 
Figure \ref{fig:time_evo_disk_after} shows the time evolution of $\Sigma_{\rm g}$ and $N_{i, \rm vap}'$ as a function of the time $t - t_{\rm p}$ after core insertion.
The NH$_3$ ice and ammonium salt lines are located outside and inside the core orbit, respectively.
Although there is a small supply of NH$_3$ vapor after the core insertion, the amount of nitrogen vapor that accretes on the planet is largely determined by the amount of vapor that has reached the outer core orbit prior to the core insertion.
The left panel of Figure  \ref{fig:planetary_growth} displays the evolution of the planet mass.
The planet mass reaches one Jupiter mass $M_{\rm jup}$ within 0.04 Myr after the onset of gas accretion.
This timescale is shorter than the growth time of Jupiter via runaway gas accretion estimated from numerical hydrodynamic simulations, but the difference is limited to a factor of a few, up to an order of magnitude. \citep{2009Icar..199..338L, 2016ApJ...823...48T, 2017ApJ...836..227L, 2023ASPC..534..947G}.

The right panel of Figure \ref{fig:planetary_growth} shows the nitrogen enrichment of the planet $x_{\rm N,p}$ as a function of $M_{\rm p,atm}$.
Here we plot the possible range of nitrogen abundance that was originally derived from the dissociation of the ammonium salt, sublimation of NH$_3$ ice, and ammonium salt in the core, taking into account the 10 wt\% uncertainty in the abundance of salts contained in the disk dust.
To highlight the contribution of NH$_3$ vapor to nitrogen enrichment, we also plot the nitrogen abundance of accreting gas only.
The final nitrogen abundance in the planetary atmosphere exceeds the solar abundance by a factor of 1--3.
Dust containing 20 wt\% of ammonium salts yields  $x_{\rm N,p}$ comparable to that measured by Juno for Jupiter \citep{2017Sci...356..821B}.
Salts incorporated into the core and accreted NH$_3$ vapor almost equally contribute to this degree of nitrogen enrichment; either one alone results in the enrichment degree below Juno's observations.
The nitrogen abundance decreases as the planet grows. This occurs because the contribution of core dissolution to nitrogen enrichment is already accounted for at the time of the protoplanet insertion, but it should be noted that this result does not constrain the timing of core dissolution during Jupiter's growth.

\subsection{Parameter study for nitrogen abundance}
\label{subsec:parameter_survey}
The nitrogen abundance in our model depends on the model parameters listed in Table~\ref{tb:parameter}, as they influence the amount of NH$_3$ vapor accreting onto the planet.
In this subsection, we present the results of Cases 1 to 7 to study the dependence on each parameter.

\subsubsection{Protoplanet formation time}
\label{subsec:parameter_survey_tcore}
\begin{figure*}[t]
    \begin{tabular}{cc}
    \hspace{-0.7cm}
      \begin{minipage}[t]{0.475\linewidth}
        \centering
        \includegraphics[width=\linewidth, bb=0 0 540 426]{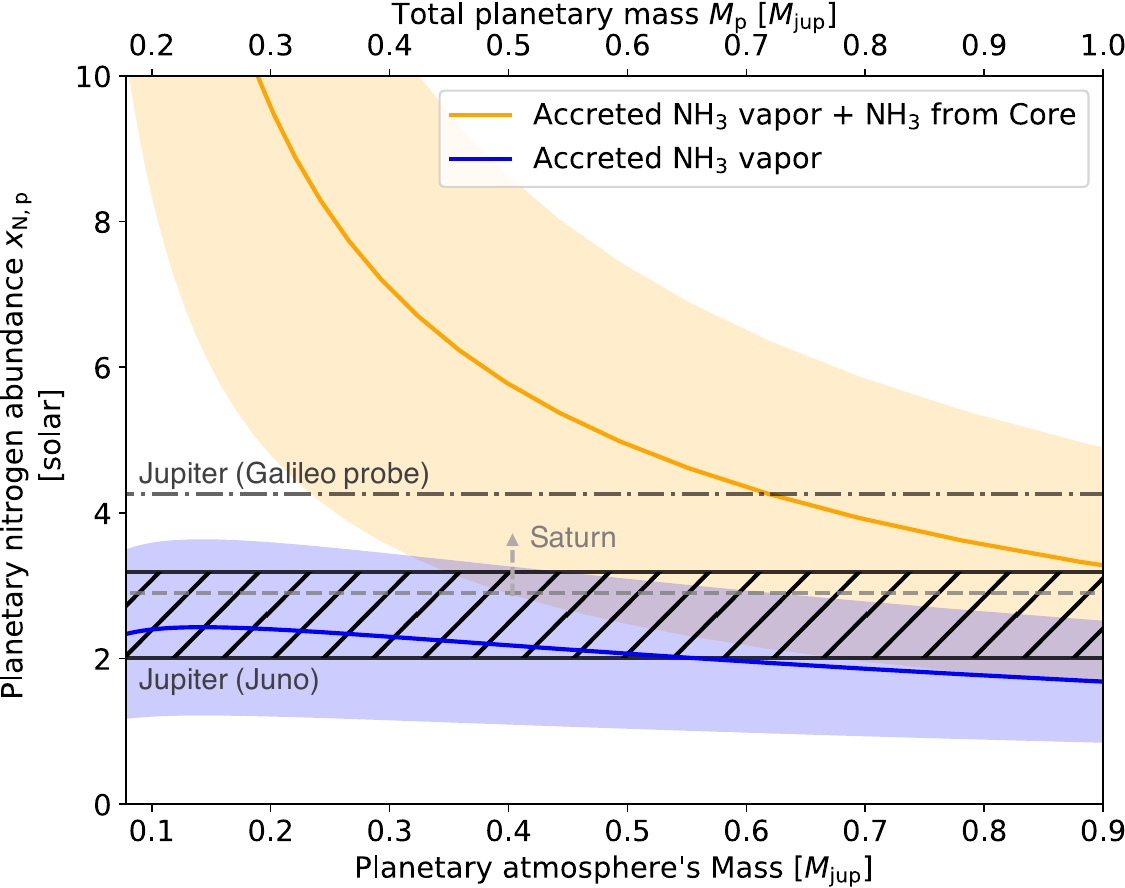}
      \end{minipage} &
        \hspace{0.5cm}
      \begin{minipage}[t]{0.475\linewidth}
        \centering
        \includegraphics[width=\linewidth, bb= 0 0 540 426]{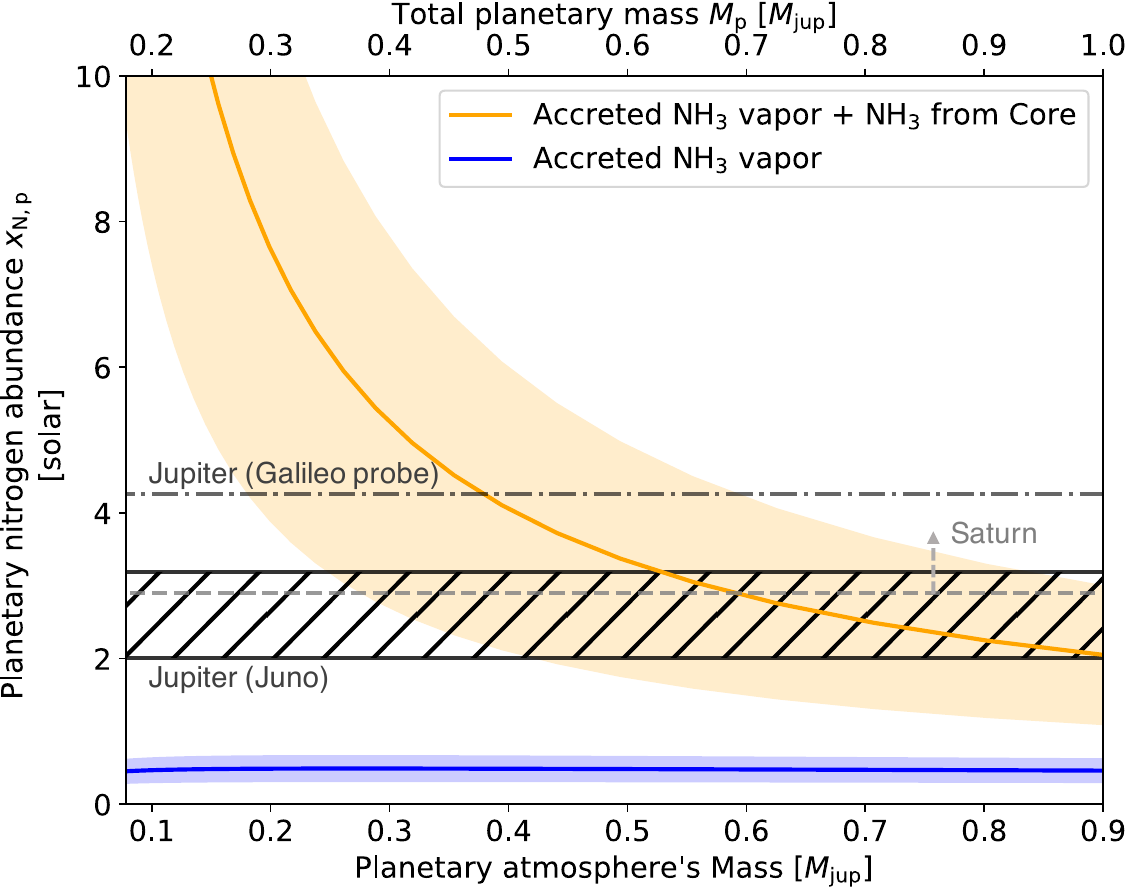}
      \end{minipage}
    \end{tabular}
    \caption{Same as the right panel of Figure \ref{fig:planetary_growth} but from the simulations with  protoplanet formation ages of 0.1 Myr (Case 1, left panel) and 1.0 Myr (Case 2, right panel). {Alt text: Two line graphs showing the evolution of atmospheric nitrogen abundance for different core insertion times.}}
\label{fig:params_tcore}
\end{figure*}

Here we present the results from Cases 1 and 2, where the protoplanet formation time $t_{\rm p}$ is changed from the fiducial value of 0.6 Myr to 0.1 and 1.0 Myr, respectively.
Figure \ref{fig:params_tcore} shows the time evolution of $x_{\rm N,p}$ for the two cases.
In case 1, the final value of $x_{\rm N,p}$ exceeds 2--5 times the solar value and is greater than that of the fiducial case because a large amount of salt-derived vapor still remains in the vicinity of the core orbit (3 au) at $t\sim$ 0.1 Myr (see Figure \ref{fig:time_evo_r2Sigma}).
This is consistent with the diffusion timescale of $\sim 0.25$ Myr around this orbit (see Section~\ref{subsubsec:result_disk_before}) being longer than $t_{\rm p}$ in Case 1.
In Case 1, the nitrogen abundance of accreted gas decreases over time because the gas accreted in the final stage of planetary growth in this case comes from large radial distances and therefore has a lower NH$_3$ abundance than that in the first stage.
Even so, if the salt comprises $>$ 20 wt\% of the dust, the salt-derived vapor alone enriches the nitrogen abundance of planetary atmospheres to a level comparable to that observed by Juno.

In contrast in Case 2, the nitrogen abundance of the accreted gas at the beginning of gas accretion is already as low as at its end because gas accretion begins after the nitrogen vapor produced by the salt and ice has diffused significantly. 
Although the NH$_3$ vapor acquired during the gas accretion phase is slightly less than in the fiducial case, the final value of $x_{\rm N,p}$ is comparable to that of the fiducial case.

\subsubsection{Core orbit}
\label{subsec:parameter_survey_rp}
\begin{figure*}[t]
    \begin{tabular}{cc}
    \hspace{-0.7cm}
      \begin{minipage}[t]{0.475\linewidth}
        \centering
        \includegraphics[width=\linewidth, bb= 0 0 540 426]{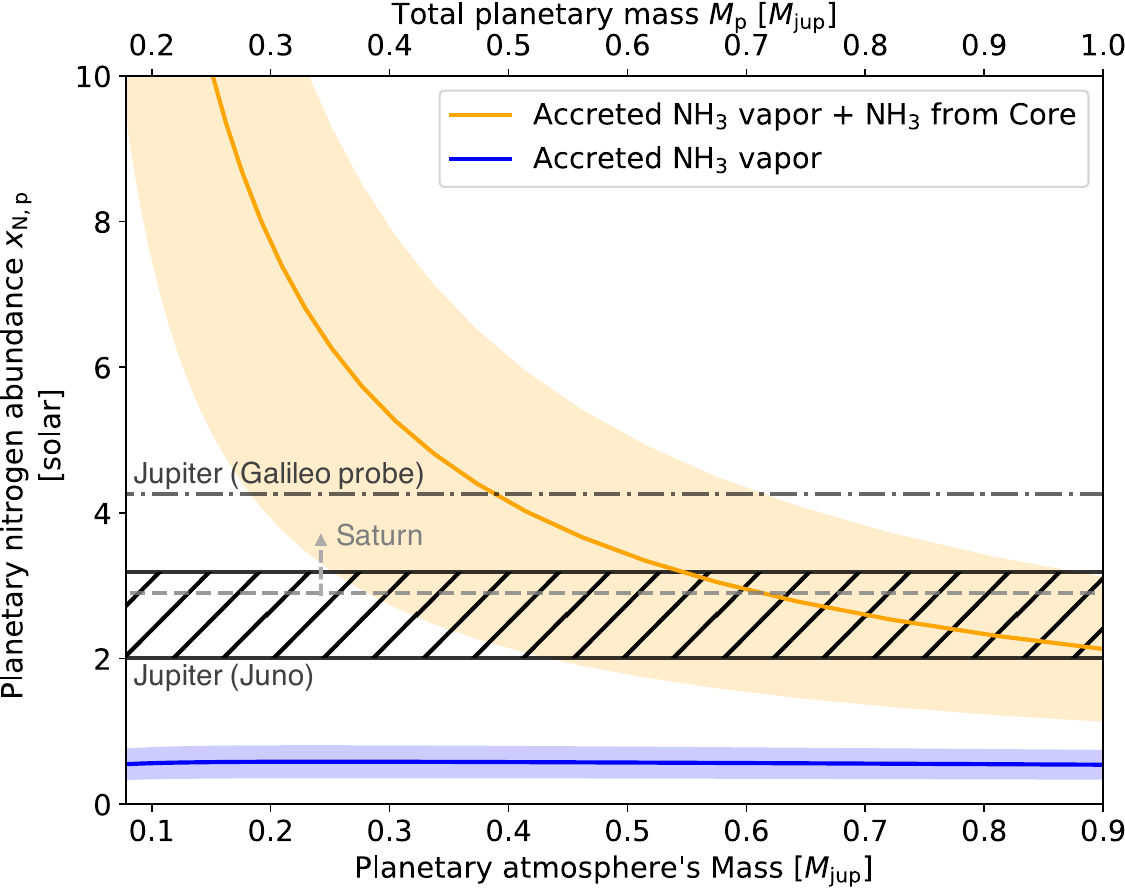}
      \end{minipage} &
        \hspace{0.5cm}
      \begin{minipage}[t]{0.475\linewidth}
        \centering
        \includegraphics[width=\linewidth, bb= 0 0 540 426]{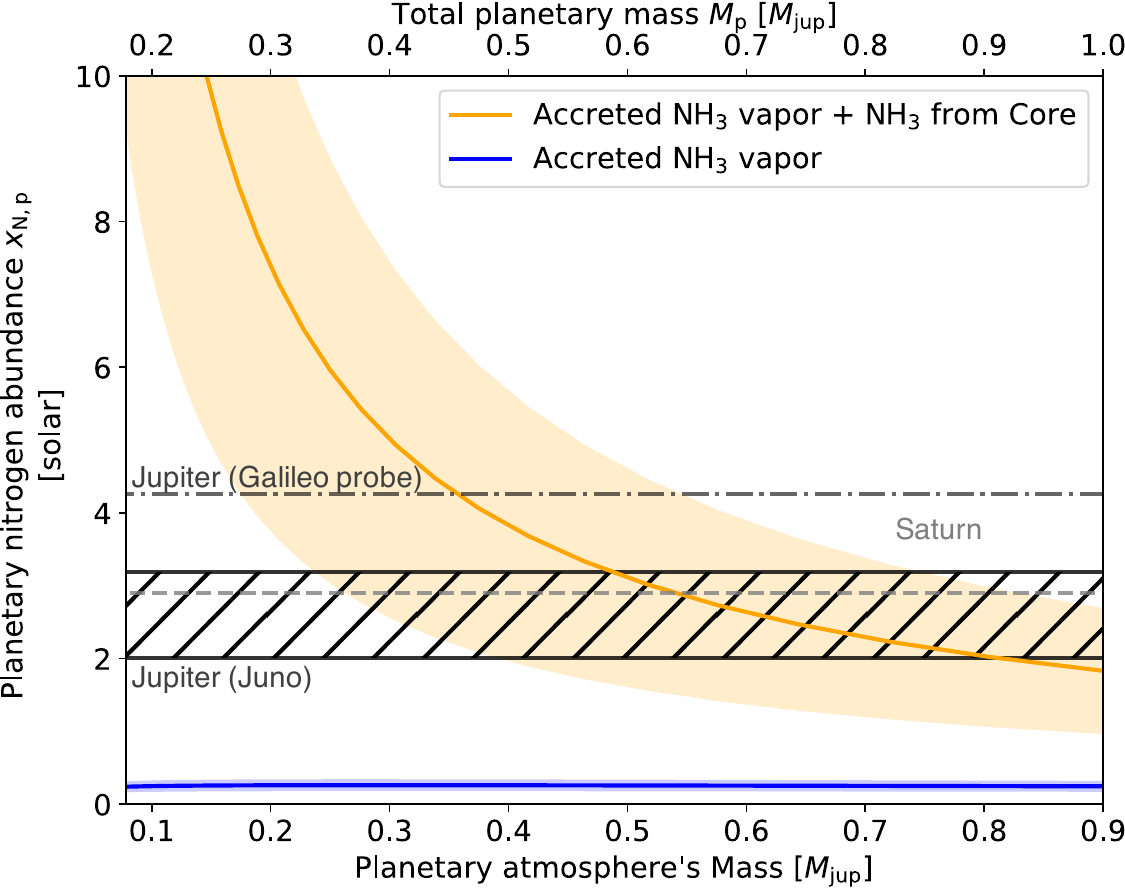}
      \end{minipage}
    \end{tabular}
    \caption{Same as the right panel of Figure \ref{fig:planetary_growth} from the simulations with  core orbits of 5.2 au (Case 3, left panel) and 10.0 au (Case 4, right panel). {Alt text: Two line graphs showing the evolution of atmospheric nitrogen abundance for different core orbits.}}
\label{fig:params_rp}
\end{figure*}

Here, we present the results from Cases 3 and 4, where the planetary orbit $r_{\rm p}$ is changed from the fiducial value of 3.0 au to 5.2 au and 10.0 au, respectively.
Figure \ref{fig:params_rp} shows the evolution of the planetary nitrogen abundance for these cases.
The final nitrogen abundance in Case 3 is 1--3 times the solar, which is comparable to that of the fiducial case. 
The final nitrogen abundance in Case 4 is  $<$ 2 times the solar even if the salt comprises 20 wt\% of the dust.
This low degrees of nitrogen enrichment are attributed to the small amounts of salt-derived NH$_3$ vapor accreting onto the distant planet.

\subsubsection{Dissociation temperature}
\label{subsec:parameter_survey_Tdis}
\begin{figure}[t]
\includegraphics[width=\hsize, bb= 0 0 540 426]{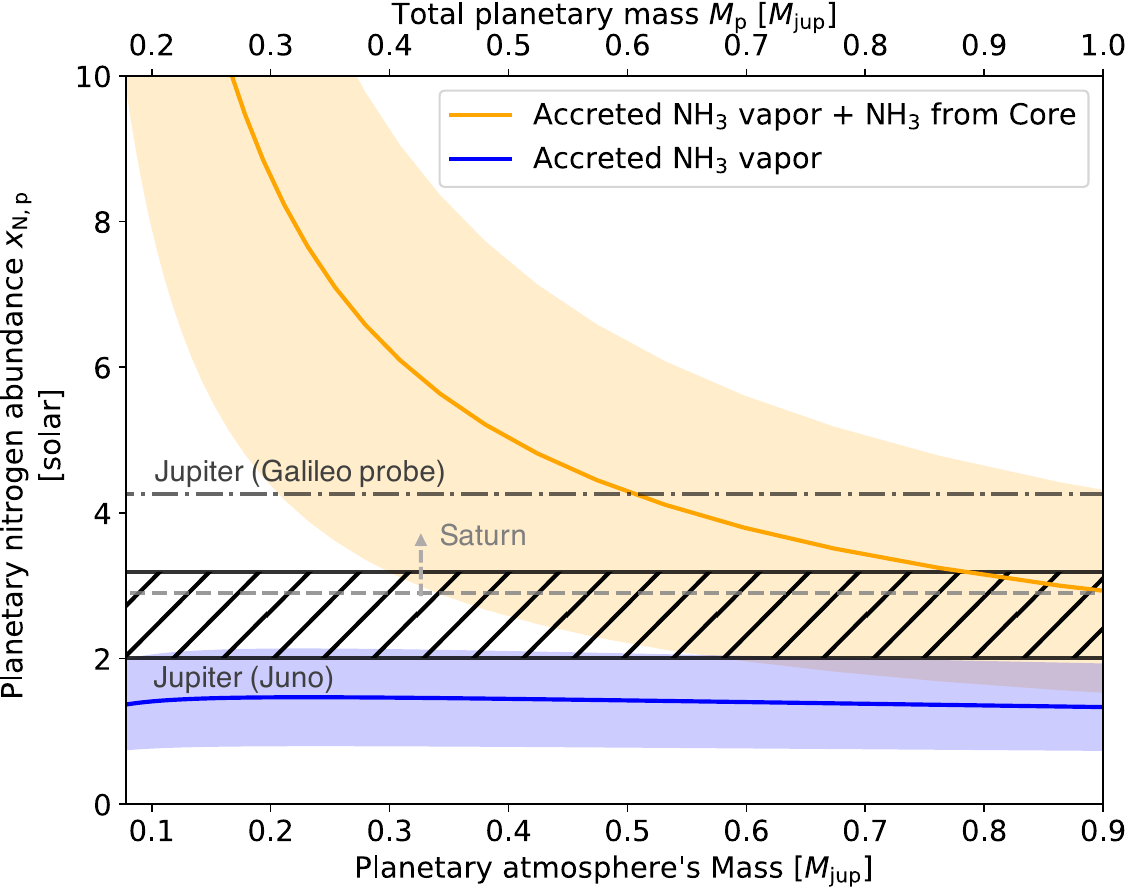}
\caption{Same as the right panel of Figure \ref{fig:planetary_growth} but from the model where the salt is assumed to be NH$_4$CN, with $T_{\rm NH_4CN, vap}=150$ K and $f_{\rm NH_3, NH_4CN} = 0.4$ (Case 5). {Alt text: Line graph showing the evolution of atmospheric nitrogen abundance when the anion of ammonium salt differs.}}
\label{fig:params_Tdiss_xp}
\end{figure}

Here we show the results from Case 5 where the salt is assumed to be NH$_4$CN.
In this case, salt dissociation occurs at a lower temperature of  $T_{\rm vap, NH_4CN} = 150$ K and hence at a large heliocentric distance of around 3 au. Therefore, salt-derived vapors are more  accessible to the core than in the fiducial case.
As shown in Figure \ref{fig:params_Tdiss_xp}, the final atmospheric nitrogen abundance in Case 5 is 2--4 times the solar, which is more than 1.5 times higher than the fiducial result.
Note that this result does not account for the nitrogen contained in the anion, HCN. If HCN is accreted by the planet, along with NH$_3$,  $x_{\rm N,p}$ would increase by a factor of 2, reaching 4--8 times the solar value.

\subsubsection{Fragmentation velocity}
\label{subsec:parameter_survey_vfrag}
\begin{figure*}[t]
    \begin{tabular}{cc}
    \hspace{-0.7cm}
      \begin{minipage}[t]{0.475\linewidth}
        \centering
        \includegraphics[width=\linewidth, bb= 0 0 532 383]{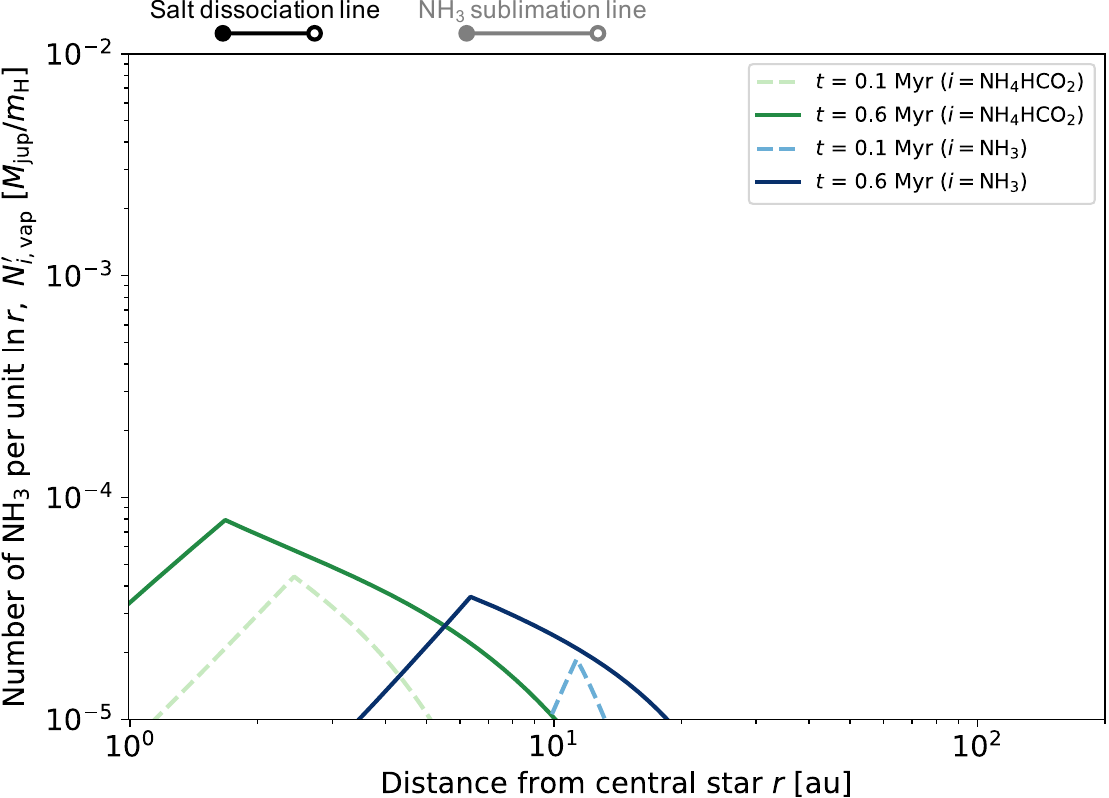}
      \end{minipage} &
        \hspace{0.5cm}
      \begin{minipage}[t]{0.475\linewidth}
        \centering
        \includegraphics[width=\linewidth, bb= 0 0 531 365]{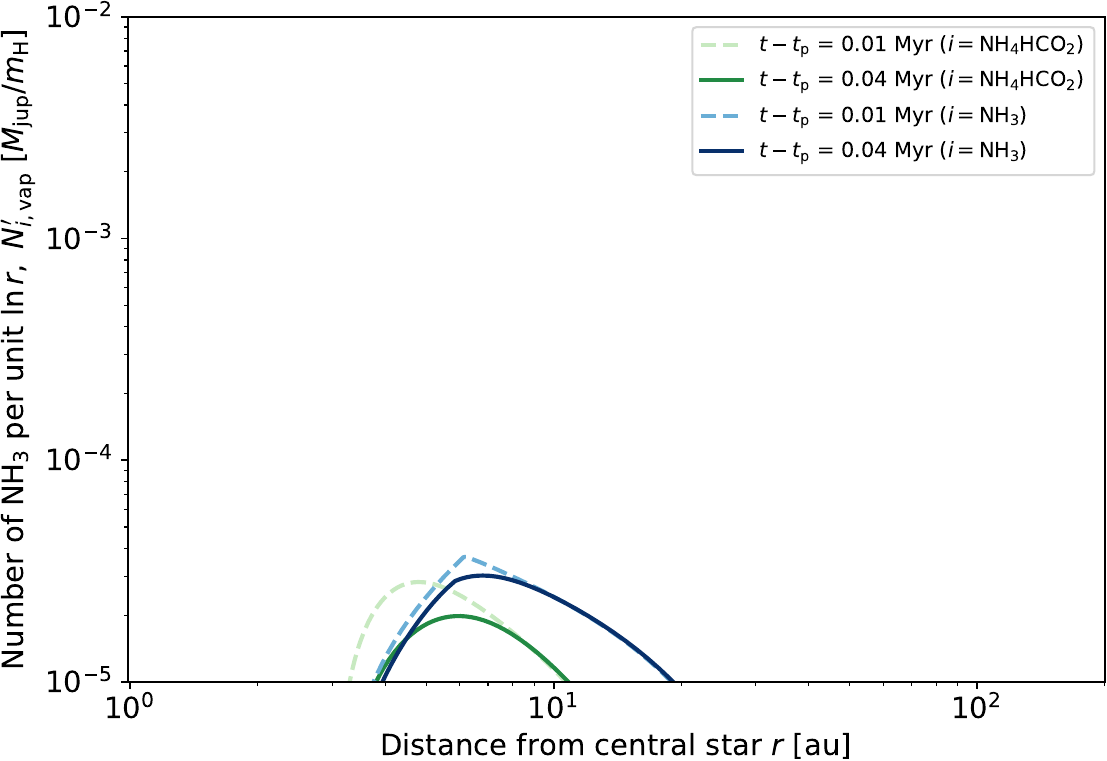}
      \end{minipage}
    \end{tabular}
    \caption{Same as Figure \ref{fig:time_evo_r2Sigma} and the right panel of Figure \ref{fig:time_evo_disk_after} but from the model with $v_{\rm frag} = 1.0~\rm m~s^{-1}$ (Case 6). {Alt text: Two line graphs showing the distribution of NH$_3$ molecules before and after core insertion for the case of low dust stickiness.}}
\label{fig:params_vfrag_r2Sigma}
\end{figure*}

\begin{figure}[t]
\includegraphics[width=\hsize, bb= 0 0 540 426]{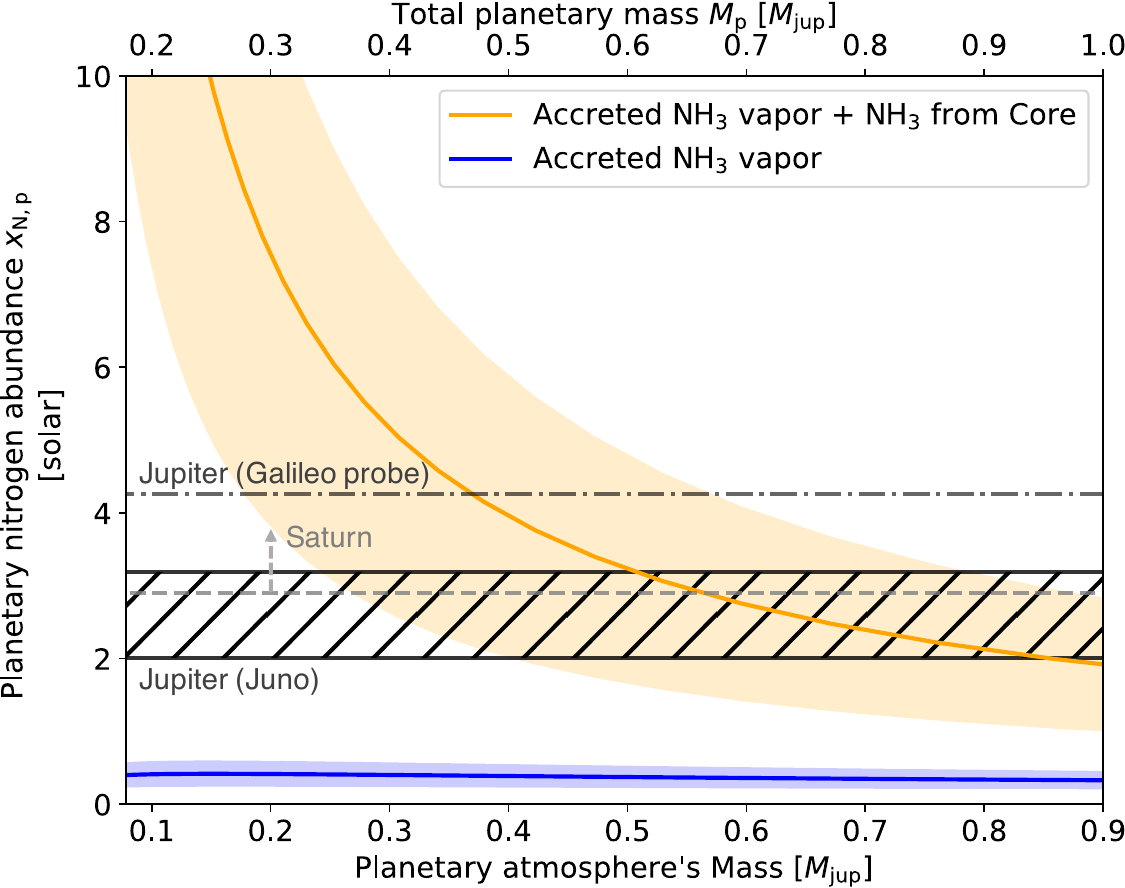}
\caption{Same as the right panel of Figure \ref{fig:planetary_growth} but from the model with $v_{\rm frag} = 1.0~\rm m~s^{-1}$ (Case 6). {Alt text: Line graph showing the evolution of atmospheric nitrogen abundance for the case of low dust stickiness.}}
\label{fig:params_vfrag_xp}
\end{figure}

To see the effect of fragmentation velocity of dust grains on nitrogen enrichment, we plot in Figures \ref{fig:params_vfrag_r2Sigma} and  \ref{fig:params_vfrag_xp} the evolution of $N_{i, \rm vap}'$ and $x_{\rm N, p}$ for Case 6, where the fragmentation velocity is taken to be 10 times lower than in the fiducial case.
The lower stickiness suppresses collisional growth and leads to smaller dust grain sizes. This, in turn, suppresses radial inward drift, resulting in a decrease in the amount of the dust passing through the NH$_3$ ice line and ammonium salt line. 
For this reason, the peaks of $N_{\rm NH_3, vap}'$ and $N_{\rm NH_4HCO_2, vap}'$ at the NH$_3$ ice line and ammonium salt line are an order of magnitude smaller than in the fiducial model.
As in Case 4, if the salt comprises 20 wt\% of the dust, the final atmospheric nitrogen abundance in Case 6 is only $\lesssim$ 2 times the solar because the amount of accreted NH$_3$ vapor is small.

\subsubsection{Turbulence strength}
\label{subsec:parameter_survey_alpha}
\begin{figure*}[t]
    \begin{tabular}{cc}
    \hspace{-0.7cm}
      \begin{minipage}[t]{0.475\linewidth}
        \centering
        \includegraphics[width=\linewidth, bb= 0 0 532 382]{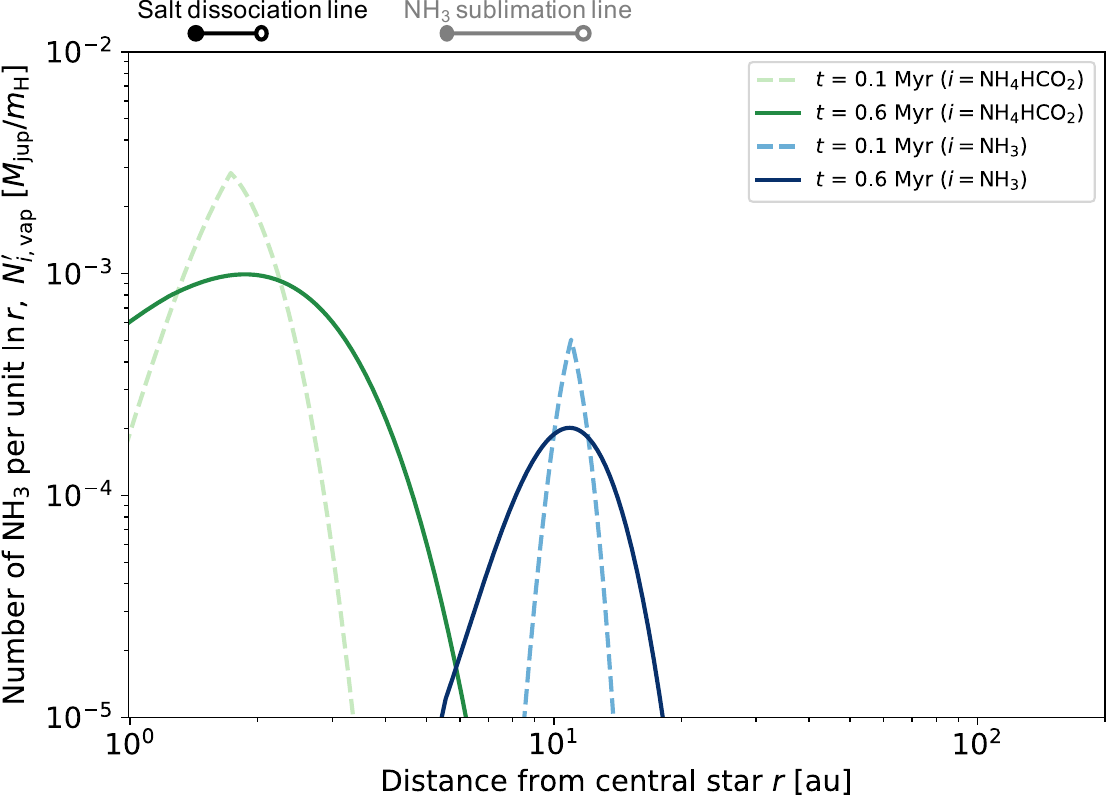}
      \end{minipage} &
        \hspace{0.5cm}
      \begin{minipage}[t]{0.475\linewidth}
        \centering
        \includegraphics[width=\linewidth, bb= 0 0 531 365]{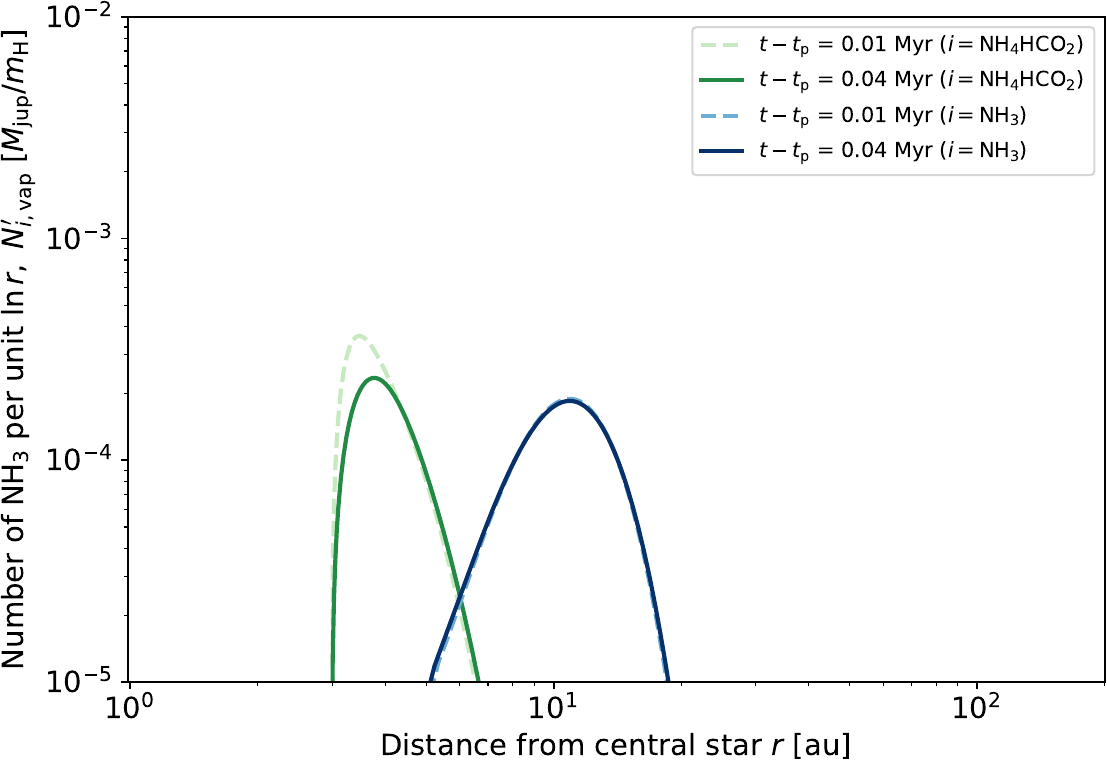}
      \end{minipage}
    \end{tabular}
    \caption{Same as Figure \ref{fig:time_evo_r2Sigma} and the right panel of Figure \ref{fig:time_evo_disk_after} but $\alpha = 10^{-4}$ (Case 7). {Alt text: Two line graphs showing the distribution of NH$_3$ molecules before and after core insertion for the case of weak turbulence.}}
\label{fig:params_alpha_r2Sigma}
\end{figure*}

\begin{figure}[t]
\includegraphics[width=\hsize, bb= 0 0 540 426]{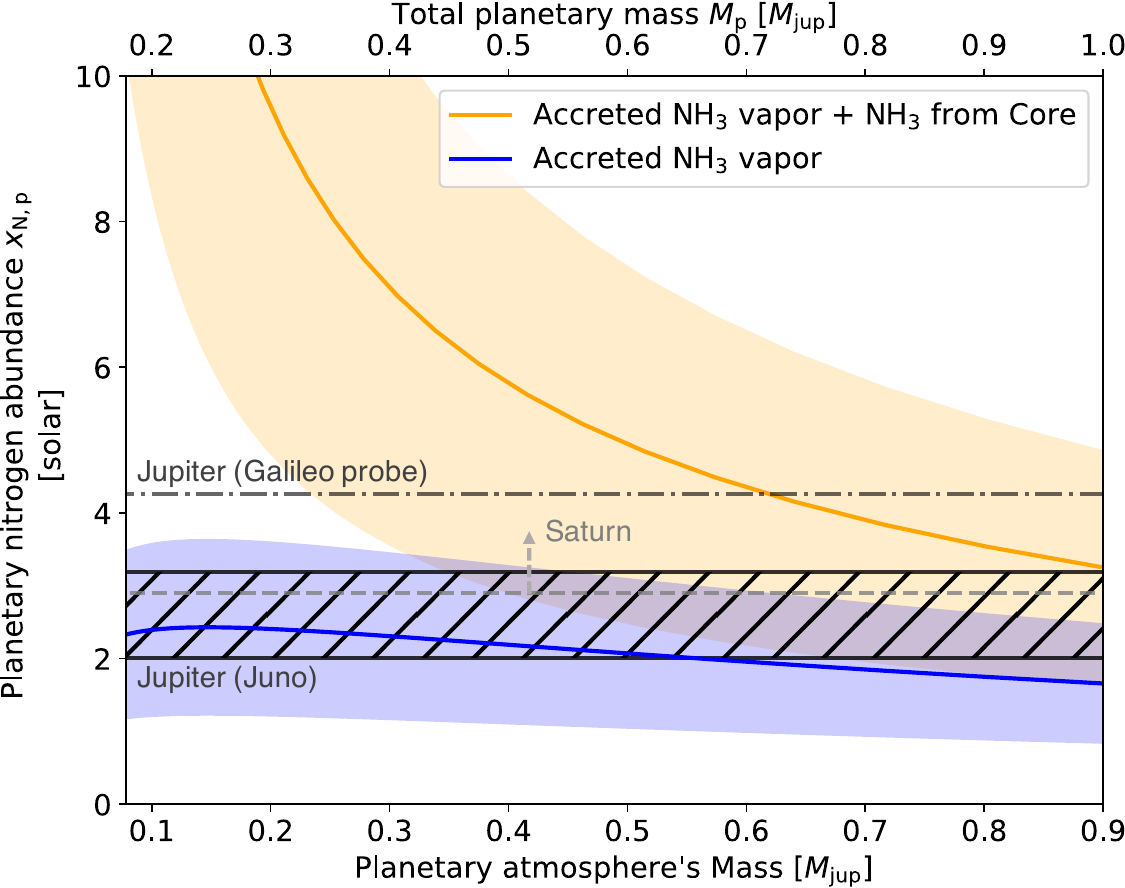}
\caption{Same as the right panel of Figure \ref{fig:planetary_growth} but $\alpha = 10^{-4}$ (Case 7). {Alt text: Line graph showing the evolution of atmospheric nitrogen abundance for the case of weak turbulence.}}
\label{fig:params_alpha_xp}
\end{figure}

Figure \ref{fig:params_alpha_r2Sigma} shows the time evolution of $N_{i, \rm vap}'$ from Case 7, where the turbulence strength $\alpha$ is 10 times smaller than in the fiducial case.
In this case, the diffusion timescale is $\sim 2$ Myr and is 10 times longer than in the fiducial model.
Thus, the diffusion of nitrogen vapor is slow, and the radial distribution of $N_{\rm NH_4HCO_2, vap}'$ and $N_{\rm NH_3, vap}'$ at the core insertion time of 0.6 My are similar to those at 0.1 Myr in the fiducial model (dotted lines in Figure \ref{fig:time_evo_r2Sigma}).
Hence, the trend of the evolution of $x_{\rm N,p}$ in Case 7 is similar to that in Case 1, where the core is inserted early (Figure \ref{fig:params_alpha_xp}). The final nitrogen enrichment degree is comparable to those in Case 1, which is higher than in the fiducial case.
Note that the weaker turbulence delays the planetary growth because the accreting gas mass on the planet depends on the gas velocity $v_{\rm g}$, which is proportional to $\alpha$ (Equation \eqref{Mp,atm}).
As a result, it takes $> 0.2$ Myr for the planet to reach Jupiter's mass.

\subsubsection{Effect of recondensation}
\label{subsec:parameter_recon}
\begin{figure*}[t]
    \begin{tabular}{cc}
    \hspace{-0.7cm}
      \begin{minipage}[t]{0.475\linewidth}
        \centering
        \includegraphics[width=\linewidth, bb= 0 0 532 384]{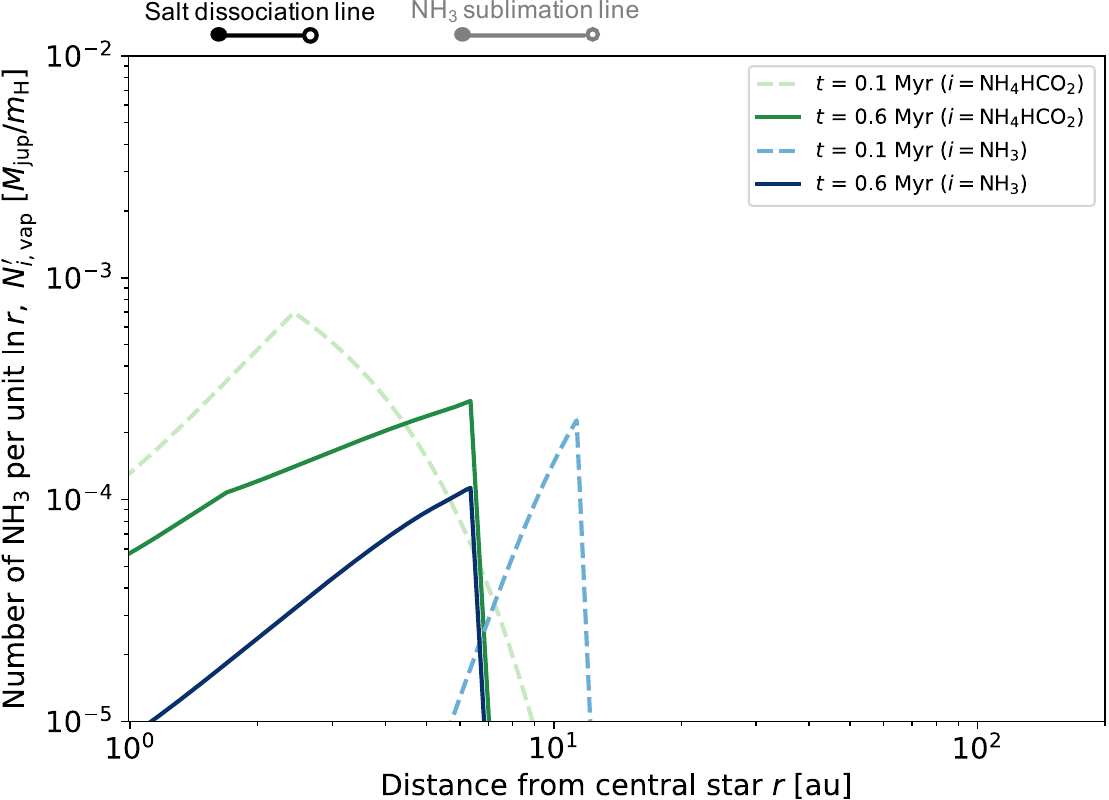}
      \end{minipage} &
        \hspace{0.5cm}
      \begin{minipage}[t]{0.475\linewidth}
        \centering
        \includegraphics[width=\linewidth, bb= 0 0 531 365]{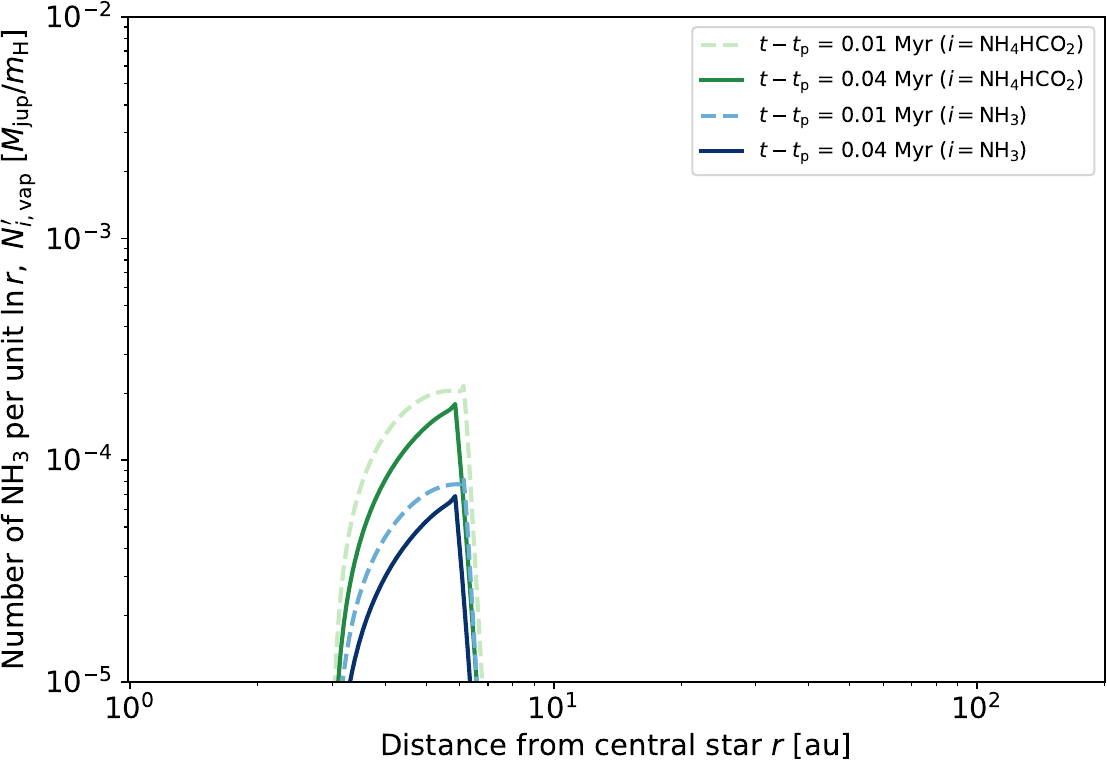}
      \end{minipage}
    \end{tabular}
    \caption{Same as Figure \ref{fig:time_evo_r2Sigma} and the right panel of Figure \ref{fig:time_evo_disk_after} but this model included the effect of recondensation (Case 8). {Alt text: Two line graphs showing the distribution of NH$_3$ molecules before and after core insertion for the model considering recondensation.}}
\label{fig:params_recon_r2Sigma}
\end{figure*}

\begin{figure}[t]
\includegraphics[width=\hsize, bb= 0 0 540 426]{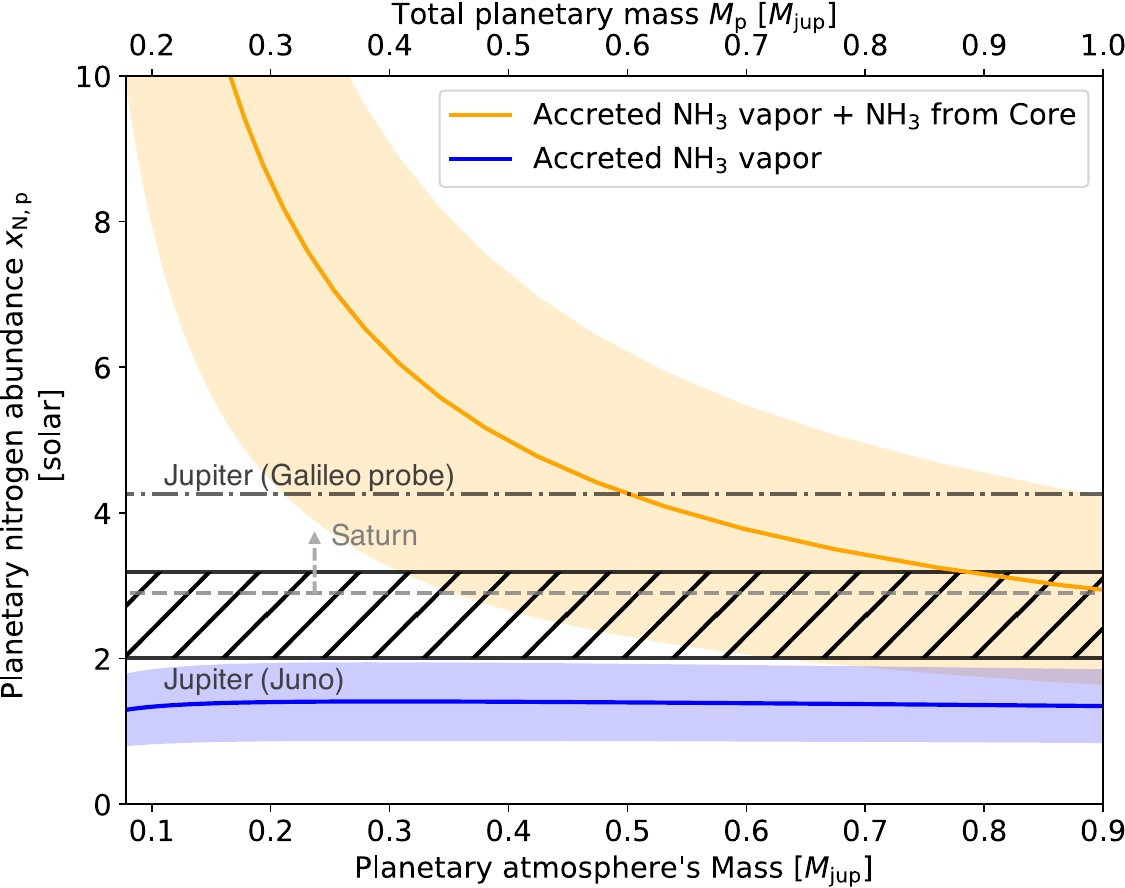}
\caption{Same as the right panel of Figure \ref{fig:planetary_growth} but this model included the effect of recondensation (Case 8). {Alt text: Line graph showing the evolution of atmospheric nitrogen abundance for the model considering recondensation.}}
\label{fig:params_recon_xp}
\end{figure}

Finally, we show the results for Case 8, where the recondensation of NH$_3$ vapors at the NH$_3$ ice line (Section \ref{subsec:model_N_species}) is introduced by equation \eqref{Sigma_i,vap_massflux_recondensation}.
Figures \ref{fig:params_recon_r2Sigma} and \ref{fig:params_recon_xp} show the evolution of $N_{i, \rm vap}'$ and $x_{\rm N,p}$ from Case 8, respectively.
Note that in Case 8, $N_{\rm NH_4HCO_2, vap}'$ and $N_{\rm NH_3, vap}'$ include the mass of the NH$_3$ ice frozen on dust outside the NH$_3$ ice line.
In this case, both salt-derived and ice-derived NH$_3$ vapors recondense onto dust as NH$_3$ ices at the outside of the NH$_3$ ice line and then drift inward.
Once the dust crosses the NH$_3$ ice line, the recondensed NH$_3$ ice reverts to vapor and stop drifting relative to the gas.
This cycle traps the NH$_3$ vapor around the NH$_3$ ice line.

However, the total nitrogen vapor outside the core orbits at the core insertion time of 0.6 Myr in Case 8 ($N'_{r > r_{\rm p}} = 0.031 M_{\rm jup}/m_{\rm H}$) is comparable to that for the fiducial case ($N'_{r > r_{\rm p}} = 0.038 M_{\rm jup}/m_{\rm H}$).
In particular, the vapor that mainly contributes to planetary growth comes from inside 5 au.
In the fiducial case, the total amount of nitrogen vapor at $r_{\rm p} < r < 5$ au is $N'_{r_{\rm p} < r < 5~{\rm au}} = 0.011 M_{\rm jup}/m_{\rm H}$, and in Case 8, $N'_{r_{\rm p} < r < 5~{\rm au}} = 0.018 M_{\rm jup}/{m_{\rm H}}$.
In Case 8, $N' = 0.0043 M_{\rm jup}/m_{\rm H}$ of NH$_3$ ice-derived vapor, which is $\sim$ 3 times grater than in the fiducial case, is present at $r < 5$ au.
The final value of $x_{\rm N,p}$ is 2--4 times the solar due in part to this vapor contribution, but it is almost comparable to that of the fiducial case.

\section{Discussion}
\label{sec:discussion}
In this section, we discuss the results presented in Section \ref{sec:results} in the context of solar system formation. Section \ref{subsec:discussion_enrichment} discusses the importance of ammonium salts as nitrogen carriers. Section \ref{subsec:discussion_oxygen_Ar} discusses the oxygen and argon enrichments in Jupiter's atmosphere, and Section \ref{subsec:discussion_planets} discusses the effects of ammonium salts on the composition of Saturn and the icy giants.
Section \ref{subsec:nitrogen_isotope} discusses the nitrogen isotopic ratios, followed by a discussion of  the potential of ammonium salts as sulfur carriers in Section \ref{subsec:discussion_sulfur}.
The influences of simplified assumptions in our model on the results are discussed in Section \ref{subsec:discussion_model}.

\subsection{Jupiter's nitrogen enhancement by ammonium salts}
\label{subsec:discussion_enrichment}
The possible presence of ammonium salts in the comet 67P \citep{2020NatAs...4..533A, 2020Sci...367.7462P} implies that the salts may have been present in the dust in the cold outer part of the solar nebula.
As demonstrated in Section~\ref{sec:results}, the salt-containing dust could have transported nitrogen to the inner disk region when they drifted inward.
If gas giants form in the inner disk ($r \sim 3$ au), salt-derived NH$_3$ can be incorporated into the planet in solid and gaseous forms during the core and gas accretion phases, respectively, contributing to nitrogen enhancement.
When the 10 $M_{\oplus}$ of heavy elements from core erosion dissolve into the planetary atmosphere, and if the dust that formed the core contained 20 wt\% ammonium salts, the atmospheric nitrogen abundance is enriched to 1.4 times the solar.
By acquiring the comparable amount of NH$_3$ vapor during gas accretion, the planet can achieve nitrogen enrichment to Jupiter's level.

Figure \ref{fig:time_evo_r2Sigma} shows that dust containing 20 \% ammonium salts can produce $10^{-4} M_{\rm jup}/m_{\rm H}$ of NH$_3$ vapor per unit $\log r$ at $r < 10$ au within 0.6 Myr.
These vapors enhance the nitrogen abundance of the planetary atmosphere forming in the vicinity of the ammonium salt line.
As a result, giant planets forming there between 0.1 and 0.6 Myr can have an atmosphere that is approximately 2--3 times more enriched in nitrogen than the Sun, a value comparable to that observed by Juno (Figures \ref{fig:planetary_growth} and \ref{fig:params_tcore}).
Similar degrees of nitrogen enrichment are attained in model parameters where salt-derived NH$_3$ vapor is accessible to the planet, such as in cases of weak turbulence or low salt dissociation temperatures (Figures \ref{fig:params_Tdiss_xp} and \ref{fig:params_alpha_xp}).
Similar enrichment also occurs even when the recondensation of NH$_3$ vapor onto dust outside the NH$_3$ snow line is taken into account (Figure \ref{fig:params_recon_xp}).
Since it is difficult to the accrete salt-derived NH$_3$ vapor in cases of core formation at large radial distance or low dust stickiness, dust containing 30 wt\% of the salt is needed to achieve Jupiter's level nitrogen enrichment (Figures \ref{fig:params_rp} and \ref{fig:params_vfrag_xp}).
On the other hand, even in the cases involving low dust stickiness and distant core formation, dust with less than 20 wt\% salt may be sufficient to enrich nitrogen to the same degree as in the fiducial case, depending on the turbulence strength and the dissociation temperature.

In contrast to the contribution of salt to the nitrogen enrichment of the gas planet, the contribution of NH$_3$ ice is small. This is because the NH$_3$ content of the dust, estimated from the comet composition, is low (0.5 wt\%), and the NH$_3$ sublimation temperature is 80 K, resulting in the formation of the NH$_3$ ice-derived vapor peak at 11 au.
If the nitrogen source of the planet is NH$_3$ ice-derived vapor alone, the nitrogen abundance of the gas planet is 0.1 times the solar in the fiducial case, and reaches a maximum value of 0.3 times the solar in case 8, where NH$_3$ ice-derived vapor is enriched near the planetary orbit due to recondensation.
These values are sub-solar and align with the conclusion by \citet{2019A&A...632L..11B} that NH$_3$ ice alone is insufficient to achieve Jupiter's nitrogen enhancement.

Our results do not account for nitrogen delivered by the refractory dust component (Section \ref{subsec:model_N_species}). \citet{2020Sci...367.7462P} suggest that organic matter may contain nitrogen at approximately 1/3 the level of salts. If nitrogen-bearing organics were dissolved into the atmosphere via core dilution, the predicted nitrogen acquisition from solid accretion could increase by up to a factor of 4/3 relative to our results.

\subsection{Enrichments of oxygen and argon}
\label{subsec:discussion_oxygen_Ar}
A notable feature of Jupiter's atmospheric composition is that a variety of elements with different volatilities, not only nitrogen, are uniformly enhanced by a factor of 2--4 relative to the protosolar value.
Particularly noteworthy is the equal enrichment of oxygen, readily accreted in solid form, and noble gases, predominantly accreted as gas, in Jupiter's orbit.
In this section, we focus on the oxygen and argon, assessing whether solid and gas accretion can collectively accounts for Jupiter's O and Ar abundances. 
We refer to Table \ref{table:abundance} in Appendix \ref{append:abundances} for Jupiter's elemental abundances.

\subsubsection{Oxygen}
\label{subsubsec:discussion_oxygen}
The O/H ratio of Jupiter's atmosphere is $\sim 3 \pm 2$ times that of the protosolar value.
The primary oxygen carriers in disks are water, silicates, and CO \citep{2021PhR...893....1O}. Since our simulations assume that Jupiter's formation occurred outside the water snowline, the amounts of H$_2$O and silicates incorporated during core accretion determine the planetary oxygen abundance.
If Jupiter's atmospheric composition reflects its bulk composition, the total oxygen mass of Jupiter is estimated to be $M_{\rm{O,jup}} \sim 3^{+2}_{-2} {\rm(O/H)}_{\rm protosolar} (m_{\rm O}/m_{\rm H}) M_{\rm jup} \sim 9^{+6}_{-6} M_{\oplus}$.
In this study, we assumed that 10 $M_{\oplus}$ of accreted dust contributed to Jupiter's atmospheric composition.
Because we adopted the rock-to-ice mass ratio of 4 for the dust grains, the planet accreted $\sim 2 M_{\oplus}$ of oxygen as water ice.
Additionally, if refractory components have an oxygen mass fraction of $\sim$ 40 wt\% akin to CI chondrites \citep{2021SSRv..217...44L}, the planet accrete up to $\sim 3 M_{\oplus}$ of oxygen as refractories.
Thus, the accreted solids onto Jupiter would compensate for 5 $M_{\oplus}$ out of the $M_{\rm{O,jup}} \sim 9 M_{\oplus}$.
If the remaining 4 $M_{\oplus}$ of oxygen were supplied by CO (entrapped by amorphous ice), CO$_2$, and ammonium salt anions, this, coupled with oxygen incorporation during solid accretion, would explain Jupiter's oxygen enrichment.
Investigating the origin of Jupiter's oxygen enrichment via disk simulations incorporating various oxygen carriers is a future work.

\subsubsection{Argon}
\label{subsubsec:discussion_argon}
Argon and other noble gases are hyper-volatile, similar to N$_2$. Noble gases are generally inert and do not form compounds.
Therefore, a mechanism such as elemental entrapment by amorphous ice is necessary to reconcile Jupiter's formation in the inner disk with the  Ar enrichment in Jupiter's atmosphere.

\citet{2019ApJ...875....9M} showed that if all argon atoms are ideally entrapped during amorphous ice deposition, the argon abundance in a planet forming at 3 au can be enriched by a factor of $\sim$ 50 relative to the protosolar value.
Laboratory experiments have demonstrated that the entrapment efficiency of Ar by amorphous ice is higher than that of N$_2$ \citep{2007Icar..190..655B, 2023ApJ...955....5S}. 
While the ice mass fraction of the dust in our model is 5 times lower than that of \citet{2019ApJ...875....9M}, Jupiter's Ar enrichment can still be explained if only 20--30\% of the total Ar budget was entrapped during amorphous ice deposition.
Thus, Jupiter's argon enrichment can be explained by incorporating noble gas entrapment by amorphous ice into our model.

In summary, we conclude that the simultaneous enrichment of Jupiter's oxygen and argon is feasible, provided that CO and Ar are entrapped by amorphous ice, and Jupiter incorporates these along with other oxygen carriers, such as CO$_2$.

\subsection{Effects of ammonium salts on other solar system gas/ice giants}
\label{subsec:discussion_planets}
In situ observations by Cassini revealed that Saturn's atmosphere is three times as enriched in nitrogen as the solar \citep{2009Icar..202..543F}.
Ammonium salts could have contributed to the nitrogen enrichment of Saturn's atmosphere.
If Saturn's formation occurred at its current orbit of 10 au, the results obtained from Case 4 imply that the nitrogen enrichment observed in Saturn's atmosphere may be primarily attributed to ammonium salts incorporated during the solid accretion phase rather than during the gas accretion phase (Figure \ref{fig:params_rp}). 
If the nitrogen-enriched atmosphere of Saturn is representative of the planet's bulk composition, the total nitrogen mass within Saturn is estimated to be $\sim 0.3M_{\oplus}$. This amount can be attained if Saturn accreted $5M_{\oplus}$ of dust with a composition of 20 wt\% NH$_4$HCO$_2$. However, dissolution of heavy elements into its atmosphere through core dilution and convective mixing like Jupiter \citep{2022Icar..37814937H} is necessary to for the heavy elements to be observable in the atmosphere.
In contrast to the necessity of nitrogen incorporation during the solid accretion phase in the case of Saturn forming in its present orbit, accretion of salt-derived vapor can be important if Saturn were to form in the inner part of the disk.
In the case with early core formation, weak turbulence, low dissociation temperature, or NH$_3$ recondensation, an atmospheric nitrogen abundance of $\gtrsim 2$ times the solar can be attained solely through the contribution of salt-derived vapor when the planetary mass reaches Saturn's mass of 0.3 $M_{\rm jup}$ (Figures \ref{fig:params_tcore}, \ref{fig:params_alpha_xp}, \ref{fig:params_Tdiss_xp}, and \ref{fig:params_recon_xp}). In these cases, the majority of the nitrogen needed for the enrichment of Saturn's atmosphere can be acquired during the gas accretion phase.

Spectral observations of the atmospheres of the solar system ice giants, Uranus and Neptune, by the Atacama Large Millimeter/submillimeter Array and the Karl G. Jansky Very Large Array reported their nitrogen abundances of $1.4^{+0.5}_{-0.3}$ and $3.9^{+2.1}_{-3.1}$ times the solar, respectively \citep{2021PSJ.....2....3M,2021PSJ.....2..105T}.
Since the nitrogen abundances of these ice giants have large uncertainties, further observations are needed to discuss the effects of nitrogen carriers on their atmospheric composition.
However, if their atmospheres are indeed enriched in nitrogen by a factor of a few, it is unlikely that vapors derived from  ammonium salts and NH$_3$ ice were the origin of the enrichment because the ice giants are far away from the ammonium salt line and the NH$_3$ ice line, which should result in less nitrogen enrichment than in Case 4.
And even if salt is incorporated during solid accretion phase, NH$_3$ would be present in the core and mantle, because the NH$_3$ produced by salt dissociation recondenses at the orbits of ice giants.
Alternative processes for nitrogen enrichment in the ice giants' atmosphere include the accretion of molecular nitrogen near the N$_2$ snow line, which exists at $T = 30$ K \citep{2017MNRAS.469.3994B}.

\subsection{Can the ammonium salts explain the nitrogen isotopic ratios of gas giants?}
\label{subsec:nitrogen_isotope}
The isotopic composition of atmospheric molecules is key to constraining the origin of planetary atmospheres.
In situ and ground-based observations of Jupiter's atmosphere report the $^{15}$N/$^{14}$N ratio of about $2 \times 10^{-3}$ \citep{2000Icar..143..223F, 2004Icar..172...50F, 2001ApJ...553L..77O, 2014Icar..238..170F}, which is close to the solar value \citep{2011Sci...332.1533M}.
Also, ground-based observations of Saturn's atmosphere by IRTF/TEXES report the upper limit of $^{15}$N/$^{14}$N ratio of $3 \times 10^{-3}$ \citep{2014Icar..238..170F}.
These observations suggest that the gas giants have an atmosphere of solar-like $^{15}$N/$^{14}$N.
In contrast, Titan has a super-solar $^{15}$N/$^{14}$N of $6  \times  10^{-3}$ \citep{2010JGRE..11512006N}.
Similarly, the $^{15}$N/$^{14}$N ratios of comet's volatile nitrogen compounds are super-solar, with $^{15}$N/$^{14}$N $=$ 6--8$\times 10^{-3}$ \citep{2008ApJ...679L..49B, 2014ApJ...780L..17R}.

Conventionally, the solar (lowest) $^{15}$N/$^{14}$N ratio is thought to originate from the accumulation of molecular nitrogen N$_2$ that has not undergone chemical reactions. 
Chemical calculations mimicking the condition of interstellar clouds show that the reaction between N atoms and nitrogen-containing ions is likely to cause isotopic fractionation and enrich $^{15}$N in nitrogen compounds such as NH$_3$ and HCN \citep{2000MNRAS.317..563T}.
Jupiter's nitrogen enrichment scenarios that assume cryogenic environments naturally explain the nitrogen isotope ratios because Jupiter's core can directly accrete N$_2$.
If the ammonium salts in the solar nebula were enriched in $^{15}$N, they are unlikely to have been the dominant source of the nitrogen in Jupiter's atmosphere.
On the other hand, when the ammonium salts were formed from NH$_3$ and an acid, the product salts may have had a low $^{15}$N/$^{14}$N because $^{14}$NH$_3$, which is lighter and therefore has a smaller vibrational energy, would be kinetically favored.
To our knowledge, there are no isotope measurements of cometary ammonium salts.
Isotope ratio measurements of the salts in comets are needed to test our scenario that the salts delivered nitrogen to the gas giants.

\subsection{Ammonium salts as sulfur carriers}
\label{subsec:discussion_sulfur}
Ammonium salts are also a potential carrier of volatile elements other than nitrogen.
\citet{2022MNRAS.516.3900A} reported a high probability of the presence of NH$_4$SH and NH$_4$F in the comet 67P. NH$_4$SH may particularly be abundant, as it exhibits a stronger spectrum than H$_2$O during cometary dust impacts.
This suggests that the ammonium salts could have served as an important sulfur carrier in the solar nebula.
The atomic ratio of sulfur to nitrogen of the solar is 0.2 \citep{2021A&A...653A.141A}, and NH$_4$SH has a five times greater impact on sulfur enhancement than on nitrogen enhancement.
By analogy with Figure 6, if the dust in a disk contains 10 wt\% NH$_4$SH, a gas giant forming in the inner disk ($\sim 3$ au) can be enriched in sulfer by a factor of $\gtrsim 5$ times compared to the solar value.
Elemental ratios of the atmosphere of gas giants such as the N/S ratio would provide important constrains on the amount and type of the salt.

\subsection{Assumptions and limitations in the model}
\label{subsec:discussion_model}
There are several simplifications in our model. In this section we discuss potential effects of these simplifications on our calculations of the disk and planetary compositions.

\textbf{Water fraction of dust: }
We have assumed a rocky dust composition with rock-to-ice mass ratio of 4 based on observations of the comet 67P \citep{2015Sci...347a3905R}. However, the dust in the protosolar nebula may have been more ice-rich (see Section \ref{subsec:model_N_species}), with a lower salt mass fraction. 
If we adopt a rock-to-ice ratio of one, the predicted nitrogen abundance in planetary atmospheres would be reduced by half compared to our results.
Even so, nitrogen enrichment to Jupiter's level can be achieved with low dissociation temperature and weak turbulence cases (Figures  \ref{fig:params_Tdiss_xp} and \ref{fig:params_alpha_xp}).

Ice abundance in dust may also affect the oxygen abundance of the planet.
In our model, planetary oxygen is primarily acquired  in the form of solids, including ice.
For a 50 wt\% ice dust composition, Jupiter would have aquired $\sim 5 M_{\oplus}$ of oxygen from ice accretion (see Section \ref{subsec:discussion_oxygen_Ar}).
By accreting rock with 40\% oxygen by mass (mainly silicates) to obtain 2 $M_{\oplus}$ of oxygen, and acquiring the remaining 2 $M_{\oplus}$ from anions of salts, CO, and CO$_2$ gases, Jupiter's oxygen enrichment can be explained.

\textbf{Heterogeneity of composition in the Jupiter's interior: }
We assumed that NH$_3$, dissolving from the core into the atmosphere, and NH$_3$ obtained during gas accretion, would exhibit uniform distribution across the entire planet, with the exception of the compact core part, owing to convective mixing.
However, Jupiter probably possesses a dilute core that extends to approximately 50\% of its radius \citep{2022Icar..37814937H}. This implies that the nitrogen composition of Jupiter's outer layers, as measured by Juno, may not accurately reflect that of the entire planet.
Although differences arise from variations in the equations of state for hydrogen and helium, Jupiter's interior models indicate a heavy-element mass fraction ranging from a few wt\% to 10 wt\% at the planetary surface, in contrast to approximately 20--30 wt\% in the region spanning 10--50\% of the radius from the planetary center \citep[e.g.][]{2017GeoRL..44.4649W, 2018A&A...610L..14V}.
However, it is uncertain whether the hypervolatile nitrogen exhibits the same compositional gradient as other heavy elements. If there is an additional presence of nitrogen in the deep interior of Jupiter, the planet would have incorporated dust containing salts in excess of 10 $M_{\oplus}$ during solid accretion.

\textbf{Salts incorporation during solid accretion: }
Our model neglects the formation process of the protoplanet and assumes that its core was formed by the accretion of a solid with the same salt content as the disk dust. For this assumption to hold, two conditions must be satisfied: 1. the ammonium salt line must be inner side of the planetary orbit at the time of protoplanet formation, and 2. the accreting solid (pebble or planetesimal) must not have undergone salt loss due to additional heating processes.
The former condition is satisfied in the fiducial case and in all cases of the parameter study, since the ammonium salt line is located at $< 3$ au.
For the latter, salts may have been lost from the planetesimals after its formation.
\citet{2021ApJ...913L..20L} calculated the growth of dust into planetesimals and the internal thermal alteration of planetesimals, showing that planetesimals that formed early lose volatiles immediately after formation by the heating due to the decay of short-lived radioactive species such as $^{26}$Al.
If planetesimals lose salts, the protoplanet gains less NH$_3$ during solid accretion, while the abundance of NH$_3$ vapor in the Jupiter-forming region would be enhanced. If the protoplanet acquires this NH$_3$ vapor during gas accretion, the total amount of NH$_3$ gained by the planet would remain unchanged from our model.

\textbf{Core dissolution degree: }
Core dissolution plays a crucial role in explaining Jupiter's nitrogen enrichment, except in cases where a low dissociation temperature (Figure \ref{fig:params_Tdiss_xp}), weak turbulence (Figure \ref{fig:params_alpha_xp}), and the effect of recondensation are considered (Figure \ref{fig:params_recon_xp}), allowing most of Jupiter's nitrogen enrichment to be explained solely by gas accretion. In our model, we have assumed that 10 $M_{\rm \oplus}$ (0.03 $M_{\rm Jup}$) of heavy elements dissolve from the core into the atmosphere. If the mass fraction of salt in the dust is 20 wt\%, this process would result in a nitrogen abundance in Jupiter's atmosphere of $\sim$ 1.4 times the solar. Assuming the mass fraction and type of salt remain constant, the degree of nitrogen enrichment due to core dissolution is proportional to the dissolved core mass. Juno observations constrain the total heavy-element mass in Jupiter to 6--27 $M_{\rm \oplus}$, implying that core dissolution could increase the nitrogen abundance in the atmosphere by up to a factor of 3.8.

\textbf{Orbital migration: } While we have assumed a core of a fixed orbit, cores with masses greater than 10 $M_{\oplus}$ migrate inward due to the Type II migration mechanism (\citealt{1986ApJ...309..846L, 1997Icar..126..261W}; for a review, \citealt{2023ASPC..534..685P}). 
However, \citet{2020ApJ...891..143T} show that the Type II migration is inefficient, and a core in the runaway gas accretion stage will only migrate by 1 au until it reaches Jupiter mass.
An inward migration by 1 au would marginally enhance the planetary nitrogen abundance because the core would then approach the ammonium salt line, facilitating the accretion of salt-derived vapors.
Nevertheless, this orbital migration minimally influences our results that the combination of salts incorporated during core accretion and salt-derived vapors acquired during gas accretion can lead to nitrogen enrichment equivalent to that of Jupiter.

\textbf{UV irradiation: }
Efficient formation of amorphous ice and entrapment of elements at the outer disk edge requires UV irradiation to desorb ice and subsequent recondensation \citep{2015ApJ...798....9M}. The removal of gas at the disk edge caused by UV photoevaporation may induce outward gas ``decretion'' \citep{2007ApJ...671..878D, 2015ApJ...815..112K}, which is neglected in our model.
However, the mass decretion rate induced by UV photoevaporation is at most $10^{-8} M_{\rm sun}$ yr$^{-1}$ \citep{2015ApJ...815..112K}, comparable to the typical gas accretion rate of the disks \citep{2016ARA&A..54..135H}.
This gas accretion rate corresponds to the $\alpha$ parameter of order 10$^{-3}$ to 10$^{-4}$, consistent with the values adopted in our model.
In this case, inward drift dominates over outward decretion with the gas when St $> \alpha$ (see Section \ref{subsec:model_dust} and footnote \ref{vd_justification} therein).
The lower left panel of Figure \ref{fig:time_evo_disk} shows that St $> \alpha$ is indeed satisfied throughout the disk.
Thus, we can conclude that the effect of photoevaporation-induced decretion on the inward dust transport would be negligible.

\textbf{Chemical reactions: }
Our model neglects the NH$_3$ or acid chemistry released with the dissociation of salts. 
The elemental abundance of the disk remains unchanged through the chemical reactions involving the molecules produced by the dissociation of salts. However, if these chemical reactions enhance the production of chemical species not typically found in the disk, they may offer clues for identifying the type and quantity of salts.
For example, the Jupiter formation scenario in the shadowed region \citep{2021A&A...651L...2O} predicts that the low-temperature shadowed region produces saturated hydrocarbons and organic molecules such as NH$_2$CHO \citep{2022ApJ...936..188N}.
In our scenario, in contrast, salt dissociation around the ammonium salt line enhances the abundances of NH$_3$ and acid vapors.
Identifying the dominant chemical products resulting from the dissociation of ammonium salts is a topic for future investigation.

\section{Summary and Conclusions}
\label{sec:conclusion}
The origin of Jupiter's nitrogen enhancement is the key to constraining the Jupiter formation process.
Ammonium salts, recently detected in the comet 67P, may have been the primary nitrogen carriers in primordial dust.
We have simulated the radial transport and dissociation of ammonium salts carried by dust in a protoplanetary disk, followed by the accretion of the gas and NH$_3$ vapor by a protoplanet, as well as the delivery of nitrogen to the planetary atmosphere from the salt-containing planetary core that undergoes dilution.
Our key findings are summarized as follows.
\begin{enumerate}
    \item  Dust containing 20 wt\% ammonium salts can produce $10^{-4} M_{\rm jup}/m_{\rm H}$ of NH$_3$ vapor per unit $\log r$ at $r < 10$ au within 0.6 Myr (Figure \ref{fig:time_evo_r2Sigma}).
    This salt-derived NH$_3$ vapor and ammonium salts incorporated during core accretion enhance the nitrogen abundance of the planetary atmosphere forming in the inner region.
    Gas giants forming there between 0.1 and 0.6 Myr can have an atmosphere that is approximately 2--3 times more enriched in nitrogen than the Sun, a value comparable to that observed by Juno (Figures \ref{fig:planetary_growth} and \ref{fig:params_tcore}).
    
    \item Comparable nitrogen enrichment to that of Jupiter is achieved across a wide range of model parameters.
    Particularly in the cases of early protoplanet formation, low salt dissociation temperature,and weak turbulence, the accretion of salt-derived NH$_3$ vapor alone leads to nitrogen abundance of $\gtrsim 2$ times the solar (Figures \ref{fig:params_tcore}, \ref{fig:params_Tdiss_xp}, and \ref{fig:params_alpha_xp}). A noteworthy contribution from salt-derived vapor is also found in the model incorporating the recondensation of NH$_3$ vapor outside the NH$_3$ ice line (Figure \ref{fig:params_recon_xp}).

    \item In addition to nitrogen enrichment by ammonium salts, enrichment of oxygen, argon, and sulfur could also be achieved.
    Oxygen enrichment can be achieved by acquiring 4 $M_{\oplus}$ of oxygen from CO and CO$_2$ gases and salt anions, in addition to H$_2$O ice and silicates (Section \ref{subsubsec:discussion_oxygen}).
    For argon to be transported to Jupiter's orbit, it must be entrapped by amorphous ice. Our model assumes a rocky dust with a rock-to-ice ratio of 4, but even so, previous laboratory experiments and disk simulations have shown that Jupiter's argon enrichment is achievable (Section \ref{subsubsec:discussion_argon}).
    Sulfur enrichment may occur due to salt anions. If the dust in the disk contains 10 wt\% NH$_4$SH, Jupiter's sulfur abundance can be enriched by a factor of 5 compared to the solar value (Section \ref{subsec:discussion_sulfur}).  
    
    \item Ammonium salts could have contributed to the nitrogen enrichment of Saturn's atmosphere. 
    If Saturn's formation occurred at its current orbit of 10 au, ammonium salts incorporated during the solid accretion would mainly contribute the Saturn's nitrogen enrichment (Figure \ref{fig:params_rp}).
    To account for the total nitrogen mass in Saturn, as estimated from the nitrogen abundance of Saturn's atmosphere, the planet would have to accrete 5 $M_{\oplus}$ of dust containing 20 wt\% of NH$_4$HCO$_2$.
\end{enumerate}

Our results show that ammonium salts may have contributed significantly to nitrogen transport in the disk.
Measurements of isotopic ratios of ammonium salts in comets and elemental ratios such as N/S in the inner disk region, and theoretical calculations of chemical reactions associated with salt dissociation would reveal whether salt dissociation has occurred and if the products of salt dissociation have been transported to the gas giants.

\begin{ack}
We would like to thank Kenji Furuya, Kazumasa Ohno, and Shota Notsu for useful discussions. We also thank Kazuaki Homma for advice on our numerical simulations.
We are grateful to the anonymous referee for helpful comments.
This work was supported by JSPS KAKENHI Grant Numbers JP20H00205, JP23H01227, JP23H00143, and JP23KJ0907.
\end{ack}

\bibliographystyle{apj}
\bibliography{article}

\begin{thebibliography}{}
\expandafter\ifx\csname natexlab\endcsname\relax\def\natexlab#1{#1}\fi

\bibitem[{{Adachi} {et~al.}(1976){Adachi}, {Hayashi}, \&
  {Nakazawa}}]{1976PThPh..56.1756A}
{Adachi}, I., {Hayashi}, C., \& {Nakazawa}, K. 1976, Progress of Theoretical
  Physics, 56, 1756

\bibitem[{{Altwegg} {et~al.}(2020){Altwegg}, {Balsiger}, {H{\"a}nni}, {Rubin},
  {Schuhmann}, {Schroeder}, {S{\'e}mon}, {Wampfler}, {Berthelier}, {Briois},
  {Combi}, {Gombosi}, {Cottin}, {De Keyser}, {Dhooghe}, {Fiethe}, \&
  {Fuselier}}]{2020NatAs...4..533A}
{Altwegg}, K., {Balsiger}, H., {H{\"a}nni}, N., {et~al.} 2020, Nature
  Astronomy, 4, 533

\bibitem[{{Altwegg} {et~al.}(2022){Altwegg}, {Combi}, {Fuselier}, {H{\"a}nni},
  {De Keyser}, {Mahjoub}, {M{\"u}ller}, {Pestoni}, {Rubin}, \&
  {Wampfler}}]{2022MNRAS.516.3900A}
{Altwegg}, K., {Combi}, M., {Fuselier}, S.~A., {et~al.} 2022, \mnras, 516, 3900

\bibitem[{{Anderson} {et~al.}(2023){Anderson}, {Rousselot}, {Noyelles},
  {Jehin}, \& {Mousis}}]{2023MNRAS.524.5182A}
{Anderson}, S.~E., {Rousselot}, P., {Noyelles}, B., {Jehin}, E., \& {Mousis},
  O. 2023, \mnras, 524, 5182

\bibitem[{{Asplund} {et~al.}(2021){Asplund}, {Amarsi}, \&
  {Grevesse}}]{2021A&A...653A.141A}
{Asplund}, M., {Amarsi}, A.~M., \& {Grevesse}, N. 2021, \aap, 653, A141

\bibitem[{{Atreya} {et~al.}(2020){Atreya}, {Hofstadter}, {In}, {Mousis}, {Reh},
  \& {Wong}}]{2020SSRv..216...18A}
{Atreya}, S.~K., {Hofstadter}, M.~H., {In}, J.~H., {et~al.} 2020, \ssr, 216, 18

\bibitem[{{Bar-Nun} {et~al.}(1985){Bar-Nun}, {Herman}, {Laufer}, \&
  {Rappaport}}]{1985Icar...63..317B}
{Bar-Nun}, A., {Herman}, G., {Laufer}, D., \& {Rappaport}, M.~L. 1985, \icarus,
  63, 317

\bibitem[{{Bar-Nun} {et~al.}(1988){Bar-Nun}, {Kleinfeld}, \&
  {Kochavi}}]{1988PhRvB..38.7749B}
{Bar-Nun}, A., {Kleinfeld}, I., \& {Kochavi}, E. 1988, \prb, 38, 7749

\bibitem[{{Bar-Nun} {et~al.}(2007){Bar-Nun}, {Notesco}, \&
  {Owen}}]{2007Icar..190..655B}
{Bar-Nun}, A., {Notesco}, G., \& {Owen}, T. 2007, \icarus, 190, 655

\bibitem[{{Bergner} {et~al.}(2016){Bergner}, {{\"O}berg}, {Rajappan}, \&
  {Fayolle}}]{2016ApJ...829...85B}
{Bergner}, J.~B., {{\"O}berg}, K.~I., {Rajappan}, M., \& {Fayolle}, E.~C. 2016,
  \apj, 829, 85

\bibitem[{{Bockel{\'e}e-Morvan} {et~al.}(2008){Bockel{\'e}e-Morvan}, {Biver},
  {Jehin}, {Cochran}, {Wiesemeyer}, {Manfroid}, {Hutsem{\'e}kers}, {Arpigny},
  {Boissier}, {Cochran}, {Colom}, {Crovisier}, {Milutinovic}, {Moreno},
  {Prochaska}, {Ramirez}, {Schulz}, \& {Zucconi}}]{2008ApJ...679L..49B}
{Bockel{\'e}e-Morvan}, D., {Biver}, N., {Jehin}, E., {et~al.} 2008, \apjl, 679,
  L49

\bibitem[{{Bolton} {et~al.}(2017){Bolton}, {Adriani}, {Adumitroaie}, {Allison},
  {Anderson}, {Atreya}, {Bloxham}, {Brown}, {Connerney}, {DeJong}, {Folkner},
  {Gautier}, {Grassi}, {Gulkis}, {Guillot}, {Hansen}, {Hubbard}, {Iess},
  {Ingersoll}, {Janssen}, {Jorgensen}, {Kaspi}, {Levin}, {Li}, {Lunine},
  {Miguel}, {Mura}, {Orton}, {Owen}, {Ravine}, {Smith}, {Steffes}, {Stone},
  {Stevenson}, {Thorne}, {Waite}, {Durante}, {Ebert}, {Greathouse}, {Hue},
  {Parisi}, {Szalay}, \& {Wilson}}]{2017Sci...356..821B}
{Bolton}, S.~J., {Adriani}, A., {Adumitroaie}, V., {et~al.} 2017, Science, 356,
  821

\bibitem[{{Booth} {et~al.}(2017){Booth}, {Clarke}, {Madhusudhan}, \&
  {Ilee}}]{2017MNRAS.469.3994B}
{Booth}, R.~A., {Clarke}, C.~J., {Madhusudhan}, N., \& {Ilee}, J.~D. 2017,
  \mnras, 469, 3994

\bibitem[{{Booth} \& {Ilee}(2019)}]{2019MNRAS.487.3998B}
{Booth}, R.~A., \& {Ilee}, J.~D. 2019, \mnras, 487, 3998

\bibitem[{{Bosman} {et~al.}(2019){Bosman}, {Cridland}, \&
  {Miguel}}]{2019A&A...632L..11B}
{Bosman}, A.~D., {Cridland}, A.~J., \& {Miguel}, Y. 2019, \aap, 632, L11

\bibitem[{{Brouwers} {et~al.}(2018){Brouwers}, {Vazan}, \&
  {Ormel}}]{2018A&A...611A..65B}
{Brouwers}, M.~G., {Vazan}, A., \& {Ormel}, C.~W. 2018, \aap, 611, A65

\bibitem[{{Chiang} \& {Goldreich}(1997)}]{1997ApJ...490..368C}
{Chiang}, E.~I., \& {Goldreich}, P. 1997, \apj, 490, 368

\bibitem[{{Danger} {et~al.}(2011){Danger}, {Borget}, {Chomat}, {Duvernay},
  {Theul{\'e}}, {Guillemin}, {Le Sergeant D'Hendecourt}, \&
  {Chiavassa}}]{2011A&A...535A..47D}
{Danger}, G., {Borget}, F., {Chomat}, M., {et~al.} 2011, \aap, 535, A47

\bibitem[{{de Pater} {et~al.}(2023){de Pater}, {Molter}, \&
  {Moeckel}}]{2023RemoteSens..15...5P}
{de Pater}, I., {Molter}, E.~M., \& {Moeckel}, C.~M. 2023, \remote, 15, 1313

\bibitem[{{Debras} \& {Chabrier}(2019)}]{2019ApJ...872..100D}
{Debras}, F., \& {Chabrier}, G. 2019, \apj, 872, 100

\bibitem[{{Desch}(2007)}]{2007ApJ...671..878D}
{Desch}, S.~J. 2007, \apj, 671, 878

\bibitem[{{Dr{\c{a}}{\.z}kowska} \& {Alibert}(2017)}]{2017A&A...608A..92D}
{Dr{\c{a}}{\.z}kowska}, J., \& {Alibert}, Y. 2017, \aap, 608, A92

\bibitem[{{Dubrulle} {et~al.}(1995){Dubrulle}, {Morfill}, \&
  {Sterzik}}]{1995Icar..114..237D}
{Dubrulle}, B., {Morfill}, G., \& {Sterzik}, M. 1995, \icarus, 114, 237

\bibitem[{{Feiden}(2016)}]{2016A&A...593A..99F}
{Feiden}, G.~A. 2016, \aap, 593, A99

\bibitem[{{Fletcher} {et~al.}(2014){Fletcher}, {Greathouse}, {Orton}, {Irwin},
  {Mousis}, {Sinclair}, \& {Giles}}]{2014Icar..238..170F}
{Fletcher}, L.~N., {Greathouse}, T.~K., {Orton}, G.~S., {et~al.} 2014, \icarus,
  238, 170

\bibitem[{{Fletcher} {et~al.}(2009){Fletcher}, {Orton}, {Teanby}, \&
  {Irwin}}]{2009Icar..202..543F}
{Fletcher}, L.~N., {Orton}, G.~S., {Teanby}, N.~A., \& {Irwin}, P.~G.~J. 2009,
  \icarus, 202, 543

\bibitem[{{Fouchet} {et~al.}(2004){Fouchet}, {Irwin}, {Parrish}, {Calcutt},
  {Taylor}, {Nixon}, \& {Owen}}]{2004Icar..172...50F}
{Fouchet}, T., {Irwin}, P. G.~J., {Parrish}, P., {et~al.} 2004, \icarus, 172,
  50

\bibitem[{{Fouchet} {et~al.}(2000){Fouchet}, {Lellouch}, {B{\'e}zard},
  {Encrenaz}, {Drossart}, {Feuchtgruber}, \& {de Graauw}}]{2000Icar..143..223F}
{Fouchet}, T., {Lellouch}, E., {B{\'e}zard}, B., {et~al.} 2000, \icarus, 143,
  223

\bibitem[{{Gautier} {et~al.}(2001){Gautier}, {Hersant}, {Mousis}, \&
  {Lunine}}]{2001ApJ...550L.227G}
{Gautier}, D., {Hersant}, F., {Mousis}, O., \& {Lunine}, J.~I. 2001, \apjl,
  550, L227

\bibitem[{{Guillot} {et~al.}(2023){Guillot}, {Fletcher}, {Helled}, {Ikoma},
  {Line}, \& {Paramentier}}]{2023ASPC..534..947G}
{Guillot}, T., {Fletcher}, L.~N., {Helled}, R., {et~al.} 2023, in Astronomical
  Society of the Pacific Conference Series, Vol. 534, Protostars and Planets
  VII, ed. S.~{Inutsuka}, Y.~{Aikawa}, T.~{Muto}, K.~{Tomida}, \& M.~{Tamura},
  947

\bibitem[{{Guillot} {et~al.}(2020){Guillot}, {Li}, {Bolton}, {Brown},
  {Ingersoll}, {Janssen}, {Levin}, {Lunine}, {Orton}, {Steffes}, \&
  {Stevenson}}]{2020JGRE..12506404G}
{Guillot}, T., {Li}, C., {Bolton}, S.~J., {et~al.} 2020, Journal of Geophysical
  Research (Planets), 125, e06404

\bibitem[{{Hartmann} {et~al.}(1998){Hartmann}, {Calvet}, {Gullbring}, \&
  {D'Alessio}}]{1998ApJ...495..385H}
{Hartmann}, L., {Calvet}, N., {Gullbring}, E., \& {D'Alessio}, P. 1998, \apj,
  495, 385

\bibitem[{{Hartmann} {et~al.}(2016){Hartmann}, {Herczeg}, \&
  {Calvet}}]{2016ARA&A..54..135H}
{Hartmann}, L., {Herczeg}, G., \& {Calvet}, N. 2016, \araa, 54, 135

\bibitem[{{Hayashi}(1981)}]{1981PThPS..70...35H}
{Hayashi}, C. 1981, Progress of Theoretical Physics Supplement, 70, 35

\bibitem[{{Helled} {et~al.}(2022){Helled}, {Stevenson}, {Lunine}, {Bolton},
  {Nettelmann}, {Atreya}, {Guillot}, {Militzer}, {Miguel}, \&
  {Hubbard}}]{2022Icar..37814937H}
{Helled}, R., {Stevenson}, D.~J., {Lunine}, J.~I., {et~al.} 2022, \icarus, 378,
  114937

\bibitem[{{Hori} \& {Ikoma}(2011)}]{2011MNRAS.416.1419H}
{Hori}, Y., \& {Ikoma}, M. 2011, \mnras, 416, 1419

\bibitem[{{Hubeny}(1990)}]{1990ApJ...351..632H}
{Hubeny}, I. 1990, \apj, 351, 632

\bibitem[{{Iaroslavitz} \& {Podolak}(2007)}]{2007Icar..187..600I}
{Iaroslavitz}, E., \& {Podolak}, M. 2007, \icarus, 187, 600

\bibitem[{{Kalyaan} {et~al.}(2015){Kalyaan}, {Desch}, \&
  {Monga}}]{2015ApJ...815..112K}
{Kalyaan}, A., {Desch}, S.~J., \& {Monga}, N. 2015, \apj, 815, 112

\bibitem[{{Kondo} {et~al.}(2023){Kondo}, {Okuzumi}, \&
  {Mori}}]{2023ApJ...949..119K}
{Kondo}, K., {Okuzumi}, S., \& {Mori}, S. 2023, \apj, 949, 119

\bibitem[{{Kouchi}(1990)}]{1990JCrGr..99.1220K}
{Kouchi}, A. 1990, Journal of Crystal Growth, 99, 1220

\bibitem[{{Kusaka} {et~al.}(1970){Kusaka}, {Nakano}, \&
  {Hayashi}}]{1970PThPh..44.1580K}
{Kusaka}, T., {Nakano}, T., \& {Hayashi}, C. 1970, Progress of Theoretical
  Physics, 44, 1580

\bibitem[{{Lambrechts} {et~al.}(2014){Lambrechts}, {Johansen}, \&
  {Morbidelli}}]{2014A&A...572A..35L}
{Lambrechts}, M., {Johansen}, A., \& {Morbidelli}, A. 2014, \aap, 572, A35

\bibitem[{{Li} {et~al.}(2020){Li}, {Ingersoll}, {Bolton}, {Levin}, {Janssen},
  {Atreya}, {Lunine}, {Steffes}, {Brown}, {Guillot}, {Allison}, {Arballo},
  {Bellotti}, {Adumitroaie}, {Gulkis}, {Hodges}, {Li}, {Misra}, {Orton},
  {Oyafuso}, {Santos-Costa}, {Waite}, \& {Zhang}}]{2020NatAs...4..609L}
{Li}, C., {Ingersoll}, A., {Bolton}, S., {et~al.} 2020, Nature Astronomy, 4,
  609

\bibitem[{{Lichtenberg} \& {Krijt}(2021)}]{2021ApJ...913L..20L}
{Lichtenberg}, T., \& {Krijt}, S. 2021, \apjl, 913, L20

\bibitem[{{Lin} \& {Papaloizou}(1986)}]{1986ApJ...309..846L}
{Lin}, D.~N.~C., \& {Papaloizou}, J. 1986, \apj, 309, 846

\bibitem[{{Lissauer} {et~al.}(2009){Lissauer}, {Hubickyj}, {D'Angelo}, \&
  {Bodenheimer}}]{2009Icar..199..338L}
{Lissauer}, J.~J., {Hubickyj}, O., {D'Angelo}, G., \& {Bodenheimer}, P. 2009,
  \icarus, 199, 338

\bibitem[{{Lodders}(2021)}]{2021SSRv..217...44L}
{Lodders}, K. 2021, \ssr, 217, 44

\bibitem[{{Lozovsky} {et~al.}(2017){Lozovsky}, {Helled}, {Rosenberg}, \&
  {Bodenheimer}}]{2017ApJ...836..227L}
{Lozovsky}, M., {Helled}, R., {Rosenberg}, E.~D., \& {Bodenheimer}, P. 2017,
  \apj, 836, 227

\bibitem[{{Lynden-Bell} \& {Pringle}(1974)}]{1974MNRAS.168..603L}
{Lynden-Bell}, D., \& {Pringle}, J.~E. 1974, \mnras, 168, 603

\bibitem[{{Marty} {et~al.}(2011){Marty}, {Chaussidon}, {Wiens}, {Jurewicz}, \&
  {Burnett}}]{2011Sci...332.1533M}
{Marty}, B., {Chaussidon}, M., {Wiens}, R.~C., {Jurewicz}, A.~J.~G., \&
  {Burnett}, D.~S. 2011, Science, 332, 1533

\bibitem[{{McClure} {et~al.}(2023){McClure}, {Rocha}, {Pontoppidan}, {Crouzet},
  {Chu}, {Dartois}, {Lamberts}, {Noble}, {Pendleton}, {Perotti}, {Qasim},
  {Rachid}, {Smith}, {Sun}, {Beck}, {Boogert}, {Brown}, {Caselli}, {Charnley},
  {Cuppen}, {Dickinson}, {Drozdovskaya}, {Egami}, {Erkal}, {Fraser}, {Garrod},
  {Harsono}, {Ioppolo}, {Jim{\'e}nez-Serra}, {Jin}, {J{\o}rgensen},
  {Kristensen}, {Lis}, {McCoustra}, {McGuire}, {Melnick}, {{\~A}-berg},
  {Palumbo}, {Shimonishi}, {Sturm}, {van Dishoeck}, \&
  {Linnartz}}]{2023NatAs...7..431M}
{McClure}, M.~K., {Rocha}, W.~R.~M., {Pontoppidan}, K.~M., {et~al.} 2023,
  Nature Astronomy, 7, 431

\bibitem[{{Molter} {et~al.}(2021){Molter}, {de Pater}, {Luszcz-Cook},
  {Tollefson}, {Sault}, {Butler}, \& {de Boer}}]{2021PSJ.....2....3M}
{Molter}, E.~M., {de Pater}, I., {Luszcz-Cook}, S., {et~al.} 2021, \psj, 2, 3

\bibitem[{{Monga} \& {Desch}(2015)}]{2015ApJ...798....9M}
{Monga}, N., \& {Desch}, S. 2015, \apj, 798, 9

\bibitem[{{Mori} {et~al.}(2021){Mori}, {Okuzumi}, {Kunitomo}, \&
  {Bai}}]{2021ApJ...916...72M}
{Mori}, S., {Okuzumi}, S., {Kunitomo}, M., \& {Bai}, X.-N. 2021, \apj, 916, 72

\bibitem[{{Mousis} {et~al.}(2019){Mousis}, {Ronnet}, \&
  {Lunine}}]{2019ApJ...875....9M}
{Mousis}, O., {Ronnet}, T., \& {Lunine}, J.~I. 2019, \apj, 875, 9

\bibitem[{{M{\"u}ller} {et~al.}(2020){M{\"u}ller}, {Helled}, \&
  {Cumming}}]{2020A&A...638A.121M}
{M{\"u}ller}, S., {Helled}, R., \& {Cumming}, A. 2020, \aap, 638, A121

\bibitem[{{Mumma} \& {Charnley}(2011)}]{2011ARA&A..49..471M}
{Mumma}, M.~J., \& {Charnley}, S.~B. 2011, \araa, 49, 471

\bibitem[{{Nakamoto} \& {Nakagawa}(1994)}]{1994ApJ...421..640N}
{Nakamoto}, T., \& {Nakagawa}, Y. 1994, \apj, 421, 640

\bibitem[{{Ni}(2019)}]{2019A&A...632A..76N}
{Ni}, D. 2019, \aap, 632, A76

\bibitem[{{Niemann} {et~al.}(2010){Niemann}, {Atreya}, {Demick}, {Gautier},
  {Haberman}, {Harpold}, {Kasprzak}, {Lunine}, {Owen}, \&
  {Raulin}}]{2010JGRE..11512006N}
{Niemann}, H.~B., {Atreya}, S.~K., {Demick}, J.~E., {et~al.} 2010, Journal of
  Geophysical Research (Planets), 115, E12006

\bibitem[{{Notsu} {et~al.}(2022){Notsu}, {Ohno}, {Ueda}, {Walsh}, {Eistrup}, \&
  {Nomura}}]{2022ApJ...936..188N}
{Notsu}, S., {Ohno}, K., {Ueda}, T., {et~al.} 2022, \apj, 936, 188

\bibitem[{{{\"O}berg} \& {Bergin}(2021)}]{2021PhR...893....1O}
{{\"O}berg}, K.~I., \& {Bergin}, E.~A. 2021, \physrep, 893, 1

\bibitem[{{{\"O}berg} \& {Wordsworth}(2019)}]{2019AJ....158..194O}
{{\"O}berg}, K.~I., \& {Wordsworth}, R. 2019, \aj, 158, 194

\bibitem[{{Ohno} \& {Ueda}(2021)}]{2021A&A...651L...2O}
{Ohno}, K., \& {Ueda}, T. 2021, \aap, 651, L2

\bibitem[{{Okazaki} {et~al.}(2023){Okazaki}, {Marty}, {Busemann}, {Hashizume},
  {Gilmour}, {Meshik}, {Yada}, {Kitajima}, {Broadley}, {Byrne}, {F{\"u}ri},
  {Riebe}, {Krietsch}, {Maden}, {Ishida}, {Clay}, {Crowther}, {Fawcett},
  {Lawton}, {Pravdivtseva}, {Miura}, {Park}, {Bajo}, {Takano}, {Yamada},
  {Kawagucci}, {Matsui}, {Yamamoto}, {Righter}, {Sakai}, {Iwata}, {Shirai},
  {Sekimoto}, {Inagaki}, {Ebihara}, {Yokochi}, {Nishiizumi}, {Nagao}, {Lee},
  {Kano}, {Caffee}, {Uemura}, {Nakamura}, {Naraoka}, {Noguchi}, {Yabuta},
  {Yurimoto}, {Tachibana}, {Sawada}, {Sakamoto}, {Abe}, {Arakawa}, {Fujii},
  {Hayakawa}, {Hirata}, {Hirata}, {Honda}, {Honda}, {Hosoda}, {Iijima},
  {Ikeda}, {Ishiguro}, {Ishihara}, {Iwata}, {Kawahara}, {Kikuchi}, {Kitazato},
  {Matsumoto}, {Matsuoka}, {Michikami}, {Mimasu}, {Miura}, {Morota},
  {Nakazawa}, {Namiki}, {Noda}, {Noguchi}, {Ogawa}, {Ogawa}, {Okada},
  {Okamoto}, {Ono}, {Ozaki}, {Saiki}, {Sakatani}, {Senshu}, {Shimaki},
  {Shirai}, {Sugita}, {Takei}, {Takeuchi}, {Tanaka}, {Tatsumi}, {Terui},
  {Tsukizaki}, {Wada}, {Yamada}, {Yamada}, {Yamamoto}, {Yano}, {Yokota},
  {Yoshihara}, {Yoshikawa}, {Yoshikawa}, {Furuya}, {Hatakeda}, {Hayashi},
  {Hitomi}, {Kumagai}, {Miyazaki}, {Nakato}, {Nishimura}, {Soejima}, {Iwamae},
  {Yamamoto}, {Yogata}, {Yoshitake}, {Fukai}, {Usui}, {Connolly}, {Lauretta},
  {Watanabe}, \& {Tsuda}}]{2023Sci...379.0431O}
{Okazaki}, R., {Marty}, B., {Busemann}, H., {et~al.} 2023, Science, 379,
  abo0431

\bibitem[{{Okuzumi} \& {Hirose}(2012)}]{2012ApJ...753L...8O}
{Okuzumi}, S., \& {Hirose}, S. 2012, \apjl, 753, L8

\bibitem[{{Okuzumi} {et~al.}(2016){Okuzumi}, {Momose}, {Sirono}, {Kobayashi},
  \& {Tanaka}}]{2016ApJ...821...82O}
{Okuzumi}, S., {Momose}, M., {Sirono}, S.-i., {Kobayashi}, H., \& {Tanaka}, H.
  2016, \apj, 821, 82

\bibitem[{{Okuzumi} {et~al.}(2012){Okuzumi}, {Tanaka}, {Kobayashi}, \&
  {Wada}}]{2012ApJ...752..106O}
{Okuzumi}, S., {Tanaka}, H., {Kobayashi}, H., \& {Wada}, K. 2012, \apj, 752,
  106

\bibitem[{{Owen} {et~al.}(2001){Owen}, {Mahaffy}, {Niemann}, {Atreya}, \&
  {Wong}}]{2001ApJ...553L..77O}
{Owen}, T., {Mahaffy}, P.~R., {Niemann}, H.~B., {Atreya}, S., \& {Wong}, M.
  2001, \apjl, 553, L77

\bibitem[{{Paardekooper} {et~al.}(2023){Paardekooper}, {Dong}, {Duffell},
  {Fung}, {Masset}, {Ogilvie}, \& {Tanaka}}]{2023ASPC..534..685P}
{Paardekooper}, S., {Dong}, R., {Duffell}, P., {et~al.} 2023, in Astronomical
  Society of the Pacific Conference Series, Vol. 534, Astronomical Society of
  the Pacific Conference Series, ed. S.~{Inutsuka}, Y.~{Aikawa}, T.~{Muto},
  K.~{Tomida}, \& M.~{Tamura}, 685

\bibitem[{{Poch} {et~al.}(2020){Poch}, {Istiqomah}, {Quirico}, {Beck},
  {Schmitt}, {Theul{\'e}}, {Faure}, {Hily-Blant}, {Bonal}, {Raponi},
  {Ciarniello}, {Rousseau}, {Potin}, {Brissaud}, {Flandinet}, {Filacchione},
  {Pommerol}, {Thomas}, {Kappel}, {Mennella}, {Moroz}, {Vinogradoff}, {Arnold},
  {Erard}, {Bockel{\'e}e-Morvan}, {Leyrat}, {Capaccioni}, {De Sanctis},
  {Longobardo}, {Mancarella}, {Palomba}, \& {Tosi}}]{2020Sci...367.7462P}
{Poch}, O., {Istiqomah}, I., {Quirico}, E., {et~al.} 2020, Science, 367,
  aaw7462

\bibitem[{{Pollack} {et~al.}(1985){Pollack}, {McKay}, \&
  {Christofferson}}]{1985Icar...64..471P}
{Pollack}, J.~B., {McKay}, C.~P., \& {Christofferson}, B.~M. 1985, \icarus, 64,
  471

\bibitem[{{Pontoppidan} {et~al.}(2019){Pontoppidan}, {Salyk}, {Banzatti},
  {Blake}, {Walsh}, {Lacy}, \& {Richter}}]{2019ApJ...874...92P}
{Pontoppidan}, K.~M., {Salyk}, C., {Banzatti}, A., {et~al.} 2019, \apj, 874, 92

\bibitem[{{Pontoppidan} {et~al.}(2014){Pontoppidan}, {Salyk}, {Bergin},
  {Brittain}, {Marty}, {Mousis}, \& {{\"O}berg}}]{2014prpl.conf..363P}
{Pontoppidan}, K.~M., {Salyk}, C., {Bergin}, E.~A., {et~al.} 2014, in
  Protostars and Planets VI, ed. H.~{Beuther}, R.~S. {Klessen}, C.~P.
  {Dullemond}, \& T.~{Henning}, 363

\bibitem[{{Potapov} {et~al.}(2019){Potapov}, {Theul{\'e}}, {J{\"a}ger}, \&
  {Henning}}]{2019ApJ...878L..20P}
{Potapov}, A., {Theul{\'e}}, P., {J{\"a}ger}, C., \& {Henning}, T. 2019, \apjl,
  878, L20

\bibitem[{{Rotundi} {et~al.}(2015){Rotundi}, {Sierks}, {Della Corte}, {Fulle},
  {Gutierrez}, {Lara}, {Barbieri}, {Lamy}, {Rodrigo}, {Koschny}, {Rickman},
  {Keller}, {L{\'o}pez-Moreno}, {Accolla}, {Agarwal}, {A'Hearn}, {Altobelli},
  {Angrilli}, {Barucci}, {Bertaux}, {Bertini}, {Bodewits}, {Bussoletti},
  {Colangeli}, {Cosi}, {Cremonese}, {Crifo}, {Da Deppo}, {Davidsson}, {Debei},
  {De Cecco}, {Esposito}, {Ferrari}, {Fornasier}, {Giovane}, {Gustafson},
  {Green}, {Groussin}, {Gr{\"u}n}, {G{\"u}ttler}, {Herranz}, {Hviid}, {Ip},
  {Ivanovski}, {Jer{\'o}nimo}, {Jorda}, {Knollenberg}, {Kramm}, {K{\"u}hrt},
  {K{\"u}ppers}, {Lazzarin}, {Leese}, {L{\'o}pez-Jim{\'e}nez}, {Lucarelli},
  {Lowry}, {Marzari}, {Epifani}, {McDonnell}, {Mennella}, {Michalik}, {Molina},
  {Morales}, {Moreno}, {Mottola}, {Naletto}, {Oklay}, {Ortiz}, {Palomba},
  {Palumbo}, {Perrin}, {Rodr{\'\i}guez}, {Sabau}, {Snodgrass}, {Sordini},
  {Thomas}, {Tubiana}, {Vincent}, {Weissman}, {Wenzel}, {Zakharov}, \&
  {Zarnecki}}]{2015Sci...347a3905R}
{Rotundi}, A., {Sierks}, H., {Della Corte}, V., {et~al.} 2015, Science, 347,
  aaa3905

\bibitem[{{Rousselot} {et~al.}(2014){Rousselot}, {Pirali}, {Jehin}, {Vervloet},
  {Hutsem{\'e}kers}, {Manfroid}, {Cordier}, {Martin-Drumel}, {Gruet},
  {Arpigny}, {Decock}, \& {Mousis}}]{2014ApJ...780L..17R}
{Rousselot}, P., {Pirali}, O., {Jehin}, E., {et~al.} 2014, \apjl, 780, L17

\bibitem[{{Sato} {et~al.}(2016){Sato}, {Okuzumi}, \&
  {Ida}}]{2016A&A...589A..15S}
{Sato}, T., {Okuzumi}, S., \& {Ida}, S. 2016, \aap, 589, A15

\bibitem[{{Schneider} \& {Bitsch}(2021)}]{2021A&A...654A..71S}
{Schneider}, A.~D., \& {Bitsch}, B. 2021, \aap, 654, A71

\bibitem[{{Schoonenberg} \& {Ormel}(2017)}]{2017A&A...602A..21S}
{Schoonenberg}, D., \& {Ormel}, C.~W. 2017, \aap, 602, A21

\bibitem[{{Schwarz} \& {Bergin}(2014)}]{2014ApJ...797..113S}
{Schwarz}, K.~R., \& {Bergin}, E.~A. 2014, \apj, 797, 113

\bibitem[{{Shakura} \& {Sunyaev}(1973)}]{1973A&A....24..337S}
{Shakura}, N.~I., \& {Sunyaev}, R.~A. 1973, \aap, 500, 33

\bibitem[{{Simon} {et~al.}(2023){Simon}, {Rajappan}, \&
  {{\"O}berg}}]{2023ApJ...955....5S}
{Simon}, A., {Rajappan}, M., \& {{\"O}berg}, K.~I. 2023, \apj, 955, 5

\bibitem[{{Takeuchi} \& {Lin}(2002)}]{2002ApJ...581.1344T}
{Takeuchi}, T., \& {Lin}, D.~N.~C. 2002, \apj, 581, 1344

\bibitem[{{Tanaka} {et~al.}(2020){Tanaka}, {Murase}, \&
  {Tanigawa}}]{2020ApJ...891..143T}
{Tanaka}, H., {Murase}, K., \& {Tanigawa}, T. 2020, \apj, 891, 143

\bibitem[{{Tanigawa} \& {Tanaka}(2016)}]{2016ApJ...823...48T}
{Tanigawa}, T., \& {Tanaka}, H. 2016, \apj, 823, 48

\bibitem[{{Terzieva} \& {Herbst}(2000)}]{2000MNRAS.317..563T}
{Terzieva}, R., \& {Herbst}, E. 2000, \mnras, 317, 563

\bibitem[{{Tollefson} {et~al.}(2021){Tollefson}, {de Pater}, {Molter}, {Sault},
  {Butler}, {Luszcz-Cook}, \& {DeBoer}}]{2021PSJ.....2..105T}
{Tollefson}, J., {de Pater}, I., {Molter}, E.~M., {et~al.} 2021, \psj, 2, 105

\bibitem[{{Valletta} \& {Helled}(2020)}]{2020ApJ...900..133V}
{Valletta}, C., \& {Helled}, R. 2020, \apj, 900, 133

\bibitem[{{Vazan} {et~al.}(2018){Vazan}, {Helled}, \&
  {Guillot}}]{2018A&A...610L..14V}
{Vazan}, A., {Helled}, R., \& {Guillot}, T. 2018, \aap, 610, L14

\bibitem[{{Wada} {et~al.}(2009){Wada}, {Tanaka}, {Suyama}, {Kimura}, \&
  {Yamamoto}}]{2009ApJ...702.1490W}
{Wada}, K., {Tanaka}, H., {Suyama}, T., {Kimura}, H., \& {Yamamoto}, T. 2009,
  \apj, 702, 1490

\bibitem[{{Wahl} {et~al.}(2017){Wahl}, {Hubbard}, {Militzer}, {Guillot},
  {Miguel}, {Movshovitz}, {Kaspi}, {Helled}, {Reese}, {Galanti}, {Levin},
  {Connerney}, \& {Bolton}}]{2017GeoRL..44.4649W}
{Wahl}, S.~M., {Hubbard}, W.~B., {Militzer}, B., {et~al.} 2017, \grl, 44, 4649

\bibitem[{{Ward}(1997)}]{1997Icar..126..261W}
{Ward}, W.~R. 1997, \icarus, 126, 261

\bibitem[{{Weidenschilling}(1977)}]{1977MNRAS.180...57W}
{Weidenschilling}, S.~J. 1977, \mnras, 180, 57

\bibitem[{{Whipple}(1972)}]{1972fpp..conf..211W}
{Whipple}, F.~L. 1972, in From Plasma to Planet, ed. A.~{Elvius} (Wiley
  Interscience Division, New York), 211

\bibitem[{{Wong} {et~al.}(2004){Wong}, {Mahaffy}, {Atreya}, {Niemann}, \&
  {Owen}}]{2004Icar..171..153W}
{Wong}, M.~H., {Mahaffy}, P.~R., {Atreya}, S.~K., {Niemann}, H.~B., \& {Owen},
  T.~C. 2004, \icarus, 171, 153

\bibitem[{{Youdin} \& {Lithwick}(2007)}]{2007Icar..192..588Y}
{Youdin}, A.~N., \& {Lithwick}, Y. 2007, \icarus, 192, 588

\bibitem[{{Zhou} {et~al.}(2024){Zhou}, {Simon}, {{\"O}berg}, \&
  {Rajappan}}]{2024ApJ...972..189Z}
{Zhou}, Q., {Simon}, A., {{\"O}berg}, K.~I., \& {Rajappan}, M. 2024, \apj, 972,
  189

\end{thebibliography}

\appendix

\section{Elemental abundances in the solar system giant planets}
\label{append:abundances}

\begin{table*}[t]
\begin{center}
\tbl{Elemental abundances in the atmospheres of the solar giant planets normalized by the protosolar values}{%
\begin{tabular}{lccccc}
\hline\noalign{\vskip 3pt} 
Element & Protosolar\footnotemark[a] & Jupiter/Protosolar\footnotemark[b] & Saturn/Protosolar\footnotemark[b] & Uranus/Protosolar\footnotemark[d] & Neptune/Protosolar\footnotemark[d] \\
\hline\noalign{\vskip 3pt}
 O/H & $5.64 \times 10^{-4}$ & 2.6$^{+2.3}_{-1.6}$ \footnotemark[c] & Not available (NA) & NA & NA \\
 C/H  & $3.31 \times 10^{-4}$ & $3.60 \pm 0.88$ & $8.01\pm0.30$ & 50--80 & 50--80 \\
 N/H  & $7.78 \times 10^{-5}$ & $2.61 \pm 0.59$ (J)\footnotemark[c]&$\ge 2.91\pm0.73$ &$1.4^{+0.5}_{-0.3}$\footnotemark[e] & $3.9^{+2.1}_{-3.1}$\footnotemark[f]\\
 & & $4.27\pm 1.63$ (G)& & &\\
 S/H & $1.52 \times 10^{-5}$ & $2.93\pm 0.69$ & $12.37$ & 10--30 & 40--50\\
 Ar/H & $2.78 \times 10^{-6}$ & $3.27 \pm 0.65$ & NA & NA & NA \\
 P/H & $2.95 \times 10^{-7}$ & $3.66\pm 0.20$& $12.34\pm0.81$ & NA & NA\\
\hline\noalign{\vskip3pt}
\end{tabular}}
\label{table:abundance}
\begin{tabnote}
\par\noindent
\footnotemark[a] From Table B.1 in \citet{2021A&A...653A.141A}
\par\noindent
\footnotemark[b] From Table 1 in \citet{2020SSRv..216...18A}, except for Jupiter's O/H. See also references therein. (J) and (G) represent observational data from the Juno and Galileo probe, respectively.
\par\noindent
\footnotemark[c] From \citet{2020NatAs...4..609L}
\par\noindent
\footnotemark[d] From Sections 5 and 6 in \citet{2023RemoteSens..15...5P}. See also references therein.
\par\noindent
\footnotemark[e] From \citet{2021PSJ.....2....3M}
\par\noindent
\footnotemark[f] From \citet{2021PSJ.....2..105T}
\par\noindent
\end{tabnote}
\end{center}
\end{table*} 
In situ exploration by spacecrafts and ground-based observations have continuously updated our understanding of the composition of the gas and ice giants in the Solar system.
We summarize in Table \ref{table:abundance} the up-to-date elemental abundances in the atmospheres of the solar system giant planets normalized by the protosolar values from \citet{2021A&A...653A.141A}.
We refer to these values as a basis for comparison and discussion of our results.
Note that the abundance in Jupiter's atmosphere varies as a function of depth and latitude. The O/H and N/H ratios of Jupiter reported by \citet{2020NatAs...4..609L} are based on Juno's measurements in the equatorial region, from 0 to 4 degrees north latitude, across a pressure range of 0.7 to 30 bar. Furthermore, the abundances of NH$_3$ and H$_2$O are proposed to may be underestimated because they can condense at high altitudes and then fall as hailstones \citep{2020JGRE..12506404G,2023ASPC..534..947G}.

\section{Elements entrapment by amorphous ice}
\label{append:entrapment}
Here we review on the entrapment of elements by amorphous water ice. Amorphous water ice is the common form of H$_2$O in cryogenic environments such as interstellar space and the outer regions of protoplanetary disks. This amorphous ice can adsorb and retain gas molecules in its matrix \citep{1985Icar...63..317B}. The adsorbed molecules are subsequently released during the phase transition of the ice from amorphous to crystalline structure, which occurs at $\sim$ 140 K \citep{1990JCrGr..99.1220K,2007Icar..190..655B}.

The molecular entrapment efficiency by amorphous water ice depends on various factors, including the type of the trapped molecules (hereafter referred to as ``target''), the target/H$_2$O ratio, and the temperature during amorphous water ice deposition.
\citet{1985Icar...63..317B} have experimentally investigated the entrapment efficiency under various conditions. They demonstrated that in a few cases of low-temperature deposition ($\sim$ 25 K) with high target/H$_2$O ratios ($>$ 3), the entrapment efficiency approaches unity. However, under other conditions, the entrapment efficiency ranges from 0.01 to 0.1.
\citet{1988PhRvB..38.7749B} found that the entrapment efficiency drops significantly above 30 K.

Recently, \citet{2023ApJ...955....5S} have re-evaluated the entrapment efficiency under conditions where the amorphous matrix number to target ratio exceeds unity (i.e. H$_2$O/target $>$ 1.0), with a deposition temperature below 30 K. They used single component gases of CO, CH$_4$, N$_2$, and Ar as the targets, and reported capture efficiencies of 0.27, 0.15, 0.15, and 0.24, respectively, when the matrix/target $\sim~$ 3.
The capture efficiencies for CH$_4$ and N$_2$ are relatively low, approximately half of those for CO and Ar. 

The difference in entrapment efficiency can be larger when the target is a multi-component gas mixture, due to intermolecular competition for adsorption onto the matrix, called competitive entrapment.
\citet{2007Icar..190..655B} examined the entrapment efficiencies for a gas mixture of CO, N$_2$, and Ar, and reported that the entrapment efficiency for N$_2$ is more than an order of magnitude lower than those for CO and Ar.
\citet{2024ApJ...972..189Z} re-evaluated the effect of competitive entrapment on mixtures of multicomponent gases, including CO, CH$_4$, N$_2$, and Ar. Their experimental results confirmed that the entrapment efficiency of N$_2$ is lower by a factor of two to an order of magnitude compared to other molecules when the deposition temperature exceeds 40 K. However, they did not reproduce the decrease in N$_2$ entrapment efficiency due to competitive entrapment at lower temperatures, as reported by \citet{2007Icar..190..655B}. They suggested that factors such as the H$_2$O/target ratio and the deposition rate of amorphous ice may influence the competitive entrapment process.
The low entrapment efficiency for N$_2$ is also supported by the fact that comets'  N$_2$/CO ratio is lower than that of the protosolar nebula value \citep{2023MNRAS.524.5182A}.
\citet{2024ApJ...972..189Z} proposed that this low N$_2$/CO may originate from amorphous ice that formed at 40--50 K in the protosolar nebula.

In summary, water amorphous ice is unlikely to entrap all target molecules even when the number of H$_2$O molecules exceeds those for the targets. And the entrapment efficiency for N$_2$ is particularly low among the volatiles, especially in gas mixtures, where it is an order of magnitude lower than for other volatiles.
Furthermore, the retention of elements through amorphous ice entrapment is effective as a mechanism for transporting the target molecule inner disk beyond its original snow line, provided the target is a volatile molecule, specifically one with a sublimation temperature lower than 143 K, the amorphous-crystalline phase transition line. Therefore, this mechanism is unlikely to affect the transport of sulfur and phosphorus, whose carriers in the disk are predominantly refractory species.

\end{document}